\documentclass{aa}  

\usepackage{supertabular,booktabs}
\usepackage{txfonts}
\usepackage{graphicx}
\usepackage{longtable}
\usepackage{caption}
\usepackage{rotating}
\usepackage{pdflscape}
\usepackage{xcolor}
\usepackage{adjustbox}
\usepackage{wrapfig}
\usepackage{sidecap}
\usepackage{float}
\usepackage{threeparttable}

\AtBeginEnvironment{longtable}{\captionsetup{type=table}}

\usepackage{txfonts}
\usepackage{natbib,twoopt}
\usepackage[breaklinks=true, colorlinks=true, urlcolor=blue, linkcolor=blue, citecolor=blue]{hyperref}
\bibpunct{(}{)}{;}{a}{}{,}             
\usepackage{placeins}
\usepackage{subfigure}
\usepackage{xcolor}

\usepackage[switch]{lineno}
    \linenumbers

\begin{document} 

   \title{Compact CO emission and no evidence of radial drift}

   \subtitle{ALMA observations of the faintest planet-forming disks in Lupus}

   \author{
          G. Ricciardi\inst{1}
          \and
          F. Zagaria \inst{2}
          \and
          A. Miotello\inst{1} 
          \and
          C. F. Manara \inst{1}
          \and
          G. Rosotti \inst{3}
          \and
          L. Zallio \inst{3}
          \and
          S. Andrews \inst{4}
          \and
          R. Booth \inst{5}
          \and
          J. Carpenter \inst{6}
          \and
          I. Cleeves \inst{7}
          \and
          S. Facchini \inst{3}
          \and
          V. V. Guzmán \inst{8}
          \and
          C. Toci \inst{9,10}
          \and
          M. Vioque \inst{1}
          \and
          D. Wilner \inst{4}
          \and
          J. P. Williams \inst{11}
    }

    \institute{
         European Southern Observatory, Karl-Schwarzschild-Str. 2, 85748 Garching, Germany 
         \and
         Max-Planck-Institut für Astronomie, Königstuhl 17, 69117 Heidelberg, Germany 
         \and
         Dipartimento di Fisica, Università degli Studi di Milano, Via Celoria 16, 20133 Milano, Italy
         \and
         Center for Astrophysics, Harvard \& Smithsonian, 60 Garden Street, Cambridge, MA 02138, USA
         \and
         School of Physics and Astronomy, University of Leeds, Leeds, LS2 9JT, UK
         \and
         Joint ALMA Observatory, Av. Alonso de C\'ordova 3107, Vitacura, Santiago, Chile
         \and
         University of Virginia (UVA), 1827 University Avenue, Charlottesville, VA, USA 
         \and
         Instituto de Astrofísica, Pontificia Universidad Católica de Chile, Av. Vicuña Mackenna 4860, 7820436 Macul, Santiago, Chile
         \and
         Universidad de Sevilla, ETSI, Camino de los Descubrimientos, 41092 Sevilla
         \and
         INAF OA Arcetri, Largo Enrico Fermi, 5, 50125 Firenze FI, Italia
         \and
        Institute for Astronomy, University of Hawaii at Manoa, 2680 Woodlawn Drive, Honolulu, HI 96822, USA
    }

    \date{\today}

    \abstract
    {A large fraction of the planet-forming disks surveyed by ALMA show faint CO emission, which is commonly interpreted as an indication of severe CO depletion. However, disks can be faint for multiple reasons, including by having their emission unresolved spatially, which may result in their size being overestimated, making their flux appear faint. The limited sensitivity of previous observations prevented us from determining whether this scenario can indeed account for the observed faint CO emission for radially compact disks, hindering our understanding of disk evolution and planet formation in most of the disk population.}
    {We present new ALMA observations targeting $^{12}$CO $(J = 3-2)$ and $^{13}$CO $(J = 3-2)$ in 17 of the faintest planet-forming disks in Lupus. We aim to test the feasibility of the compact disk scenario as a plausible explanation for compact disks with faint CO isotopologue emission.}
    {Our sample contains 17 disks observed with ALMA in Band 7 at the moderate angular resolution of $0\farcs25$ ($\approx$ 20~au radius at 160~pc, median distance of the sample), approximately one order of magnitude deeper than the available archival ALMA data where $^{12}$CO and $^{13}$CO were not detected. We used line stacking techniques to enhance the signal to noise ratio and extract the CO fluxes when possible. Finally, we compared the CO line luminosities with a grid of physical-chemical models of extended and compact disks and computed the disk dust and CO sizes.}
    {We detected $^{12}$CO and $^{13}$CO emission in 10 disks, 4 disks were detected only in $^{12}$CO, and 3 disks were not detected in either of the two isotopologues.}
    {The observations indicate that some of these disks are consistent with being intrinsically compact and optically thick, in both $^{12}$CO and $^{13}$CO. This scenario offers an alternative explanation to the commonly accepted hypothesis of significant CO depletion. The derived gas radii further support this interpretation ($R_{\rm CO} \leq 40\,{\rm au}$), suggesting that a significant fraction of disks may be born intrinsically small, as also indicated by recent Class 0/I surveys. Furthermore, the resulting gas-to-dust size ratios reveal no clear signs of dust evolution, suggesting that these compact disks are not drift-dominated.}


    \maketitle

\nolinenumbers

\section{Introduction}
Thanks to its unprecedented sensitivity and angular resolution, the Atacama Large Millimeter/submillimeter Array (ALMA) has revolutionized our understanding of star and planet formation. Over the past decade, ALMA has enabled extensive surveys of protoplanetary disks across a variety of nearby star-forming regions (SFRs), such as Lupus, Taurus, Chamaeleon I, Ophiuchus, Orion, Corona Australis and many others \citep[e.g.,][]{Ansdell16, ansdell17, Ansdell18, Pascucci16, Barenfeld16, Eisner16, Long18, Cieza19, Cazzoletti19, Grant21, Carpenter25, GuerraAlvarado25}, spanning an age range between about 1 and 10 Myr. These observations, primarily conducted at moderate spatial resolution ($0\farcs25$--$0\farcs50$) and sensitivity (0.1--0.4 $M_{\oplus}$ in dust), have targeted both the dust and gas components through submillimeter continuum and CO isotopologue line emission, measuring fluxes, and sizes for hundreds of disks \citep[for a review, see][]{Miotello23,Manara23}. 

A major outcome of these surveys is that the CO isotopologue emission is systematically fainter than expected \citep[e.g.,][]{Ansdell16, Long17, Miotello17, Williams14}. The reason for this discrepancy is still highly debated. In particular, disks with faint or undetected CO emission often exhibit compact or unresolved continuum emission, thus raising the question whether the faint CO emission reflects intrinsically small disk structures or faint but extended emission. 

The most common interpretation of faint CO emission is carbon and oxygen depletion \citep[e.g.,][]{Favre13, Kama16a, Miotello17}. For relatively massive and bright disks, for which reliable gas masses could be estimated, the inferred CO depletion factors are typically moderate, of order $5-100$ \citep[e.g.,][]{Bergin13, Schwarz16, McClure16}. Recent results suggest similar depletion levels in disks where $^{12}$CO emission was already detected in previous ALMA surveys \citep{Zhang25, Trapman25_masses, Rosotti25}. However, this likely reflects a strong selection bias, as these surveys are limited to CO-bright systems. For the fainter population, where even $^{12}$CO remains undetected, the required depletion factors could be much higher (up to $\approx1000$; \citealt{Barenfeld16}), which would be difficult to reconcile with current models of carbon depletion \citep[e.g.,][]{Krijt20}.

An alternative explanation for the faintest sources is that many of these CO-faint disks are intrinsically compact rather than solely carbon-depleted. The idea that disks with faint CO fluxes may simply be radially small is not new. \cite{Barenfeld16} proposed that the lack of CO detections in roughly half of the disks detected in the continuum in Upper Scorpius could be explained if the CO emission, while optically thick, arises from a compact region with a radius $<40\,{\rm au}$. Similarly, \cite{Pietu14}, using IRAM Plateau de Bure observations of T~Tauri stars, found that disks with faint CO and continuum emission often correspond to compact structures with high surface densities in their inner regions. They also suggested that such compact disks may represent up to $25\%$ of the entire disk population. Later studies in the Lupus SFR have provided further insight. \cite{Miotello21} suggested that disks that were not detected in $^{13}$CO and with faint or not detected $^{12}$CO emission, may be effectively compact , adopting a model-based definition in which compact disks have a critical radius $R_{\rm c}\leq15\,{\rm au}$, thereby redefining the potential amount of compact disks at $\approx50\%-60\%$ of the entire disk population. More recently, \citet{Deng25} suggested that CO abundances may be relatively similar across systems in the region, thus supporting the idea that disk structure, particularly compactness, rather than extreme or selective carbon depletion, could account for faint CO emission.

A key limitation of previous studies is that most disk models and ALMA observations were not designed for compact disks. Rather, they were biased toward bright and extended sources. For example, \cite{Miotello16} created a grid of over 800 disk models using the physical-chemical code DALI to compare with CO isotopologue spatially integrated fluxes observed in the Lupus SFR. This initial grid, built under the assumptions of (1) ISM-like volatile C and O abundances, and (2) an exponentially tapered power-law gas surface density \citep{LyndenBell74,Hartmann98} linearly decreasing with radius and with critical radii $30\,{\rm au}\leq R_c\leq200\,{\rm au}$, was not well suited to modeling the faintest CO sources in the sample. Within this framework, the only way to reproduce the faint $^{13}$CO and $^{12}$CO luminosities was to assume significant volatile C depletion by a factor of $10-100$ \citep{Miotello17}. However, motivated by the growing number of faint disks observed, \cite{Miotello21} later developed a new grid of compact disk models (scale radii $\leq15\,{\rm au}$) that could reproduce the observations, without invoking extreme levels of C depletion. The latter scenario is tantalizing, given that recent results by \cite{GuerraAlvarado25} have shown that these disks are very small in the continuum. However, due to the low sensitivity of the data by \cite{Ansdell18}, it is not yet clear whether these disks being compact in CO is a valuable explanation for their faintness. 

Carbon depletion and reduced disk sizes are not mutually exclusive. A combination of these two mechanisms may provide a more realistic explanation for the faint CO fluxes observed in several SFRs, as compact disks are intrinsically fainter, thus requiring more moderate levels of volatile carbon depletion. Assessing whether radially compact disk structures represent a viable scenario is crucial because of the implications on disk evolution and planet formation. Disk sizes are key first-order diagnostics of disk evolution \citep[e.g.,][]{Tabone25}: while providing a simplified view of a complex physical process, different evolutionary models predict qualitatively distinct size trends. Indeed, viscous models generally lead to disk expansion over time, whereas MHD-wind–driven evolution tends to maintain nearly constant or even decreasing disk sizes \citep[e.g.,][]{Trapman20, Trapman22}. However, additional processes, such as external photoevaporation or dynamical interactions with companions, can further limit the disk extent \citep[e.g.,][]{Facchini16, RosottiClarke18, Rota22, Anania25}, thus complicating the picture.

At the same time, the evolution of dust particles plays a central role in shaping the CO chemistry and the resulting molecular emission in protoplanetary disks \citep[e.g.,][]{Facchini17}. Grain growth, vertical settling and radial drift alter the penetration depth of ultraviolet photons, thereby influencing the chemical pathways, gas temperature and molecular emission. Moreover, the available dust surface area regulates the balance between CO freeze-out and desorption, directly affecting the observed CO luminosity and apparent gas extent. Therefore, accurately inferring gas properties requires accounting for dust evolution. In this context, a key diagnostic tool is the gas-to-dust size ratio ($R_{\rm CO}/R_{\rm dust}$), which is a powerful tracer of the coupling between solids and gas \citep[e.g.,][]{Toci21}.

In this paper, we present the first systematic characterization of the gaseous component of CO-faint disks in Lupus, using moderate-resolution ALMA Band 7 observations. Thanks to the sensitivity of our survey, which is nearly an order of magnitude deeper than that achieved by \citet{Ansdell18}, we can evaluate, for the first time, whether these faint disks are intrinsically compact or extended with low surface brightness for the first time. In Section~\ref{section2} we describe the observations and data reduction. In Section~\ref{section3} we present the methods used to extract integrated CO fluxes and derive CO radii. In Section~\ref{section4} we report the results of our measurements and in Section~\ref{section5} we discuss their implications in the context of disk evolution and planet formation. Finally, in Section~\ref{section6} we summarize our main conclusions.

\section{Observations} \label{section2}

\subsection{Sample} \label{section2.1}
Our sample consists of 18 disks located in the Lupus SFR. The sources were selected to be potentially compact, as their continuum emission appeared unresolved or marginally resolved and their CO emission was particularly faint \citep{Ansdell16}. Specifically, such disks had not been detected in $^{13}$CO $(J=2-1)$ emission and their $^{12}$CO $(J=2-1)$ luminosity was lower than $2.5\times10^{8}\,{\rm mJy}\,{\rm km}\,{\rm s}^{-1}$ ${\rm pc}^2~(\sim2.1 \times 10^{-2} L_{\odot}$ at an average distance of 150 pc). This cut excluded the brighter and resolved disks where large CO outer radii had already been measured \citep{Sanchis21}. Furthermore, disks that were targeted by the recent AGE-PRO \citep{Zhang25,Deng25} and DECO Large Programs were excluded. Such programs are biased toward targets that had been detected in $^{12}$CO, and with brighter and more extended continuum. The selected sample was also designed to exclude multiple systems, in which disk truncation from a stellar companions might occur. For similar reasons, Lupus provides the perfect setting in which to investigate whether faint disks are intrinsically compact, as it is a low-density \citep{Comeron08, Alcala14}, and low-FUV irradiated \citep{Anania25a} star-forming region, so environmental effects, such as external truncation due to flybys and photoevaporation due to massive OB-type stars, can reasonably be neglected. Lupus is also  one of the best-characterized nearby SFRs. Its disk population has been extensively studied at (sub-)millimeter wavelengths using ALMA in Band 3, 6, and 7 \citep{Ansdell16, Ansdell18, vanTerwisga18, Sanchis20, Tazzari21, GuerraAlvarado25, Deng25}, as well as through complementary UV to IR spectroscopy using VLT/X-Shooter spectroscopy \citep{Alcala14, Alcala17, Alcala19}. Moreover, the sample was selected to span a range of stellar host masses, from about 0.05 to $1\,M_\odot$. 

\begin{figure}[t]
    \centering
    \includegraphics[width=\columnwidth]{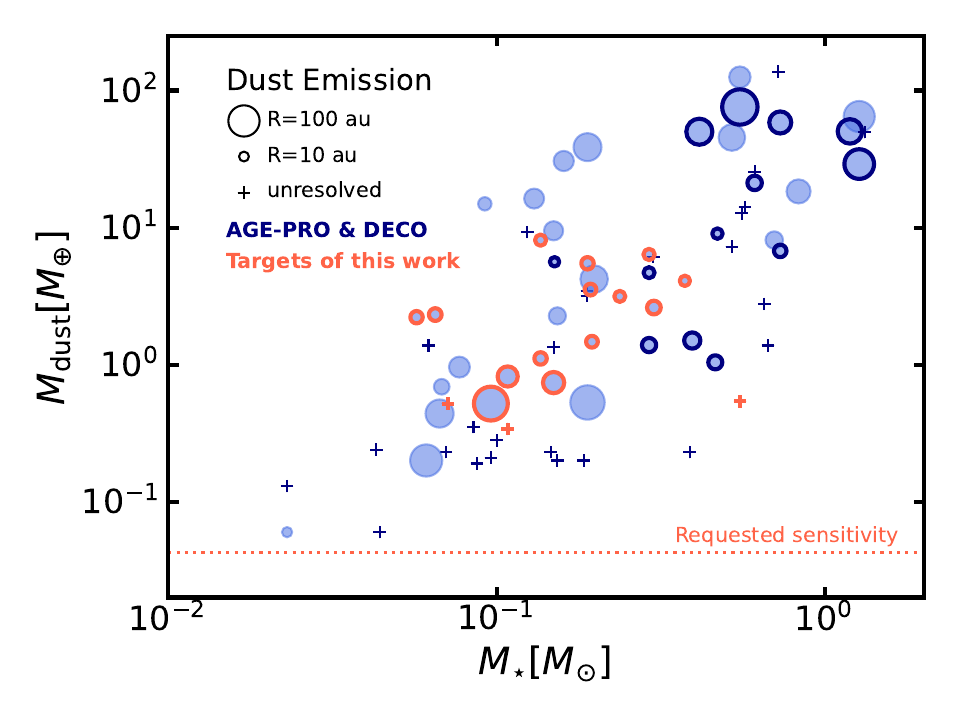}
    \caption{Entire Lupus sample. $M_{\rm dust}$, $M_{\star}$, and $R_{\rm dust}$, visualized with the symbol size, are taken from \cite{Manara23}. The disks targeted by recent ALMA Large Programs and by this work are highlighted in dark blue and red respectively. The requested continuum sensitivity of the our sample is shown with the dotted red line.}
    \label{figure:sample_M}
\end{figure}

The final sample of disks is shown in red in Fig.~\ref{figure:sample_M} (except for J15450634-3417378, for which a stellar mass measurement is not available) and their source properties are reported in Tab.~\ref{table:stellar_par}. The observations were carried out under two distinct Science Goals (SGs), each requiring different sensitivities of $9.168\,{\rm mJy}$ (corresponding to $\approx1.5\,{\rm K}$), and $1.676\,{\rm mJy}$ (corresponding to $\approx300\,{\rm mK}$), respectively. The first SG targets five sources that were already detected in $^{13}$CO $(J=3-2)$ by \cite{Ansdell16}, for which we requested a similar sensitivity to detect $^{12}$CO $(J=3-2)$, as we expected $F_{^{12}\rm CO} \geq F_{^{13}\rm CO}$. For the second set of thirteen disks, that did not have any $^{13}$CO detection in Band 7, we requested deeper observations with a sensitivity of $0.3\,{\rm K}$ over a frequency width of $2.5\,{\rm km}\,{\rm s}^{-1}$. Specifically, as shown by \cite{Ansdell18}, Sz~131, J15450634-3417378, J16081497-3857145, and J16085324-3914401 exhibit very faint emissions in $^{12}$CO and $^{13}$CO; Sz~72, J16095628-3859518, and J16134410-3736462 have only been detected in one of the two isotopologues; the rest have never been detected in CO isotopologues. J15450634-3417378 was excluded from further analysis. The presence of extended emission suggests an association with an outflow component that contaminates the CO emission and hinders reliable characterization of the disk itself.

Overall, the achieved sensitivities make this survey approximately an order of magnitude deeper than the ALMA survey of Lupus disks by \citet{Ansdell18}, enabling a significantly more detailed characterization of faint CO emission in disks that are more representative of the bulk population than most of the so-far targeted ones, as shown in Fig.~\ref{figure:sample_M}.

\subsection{Observational setup and data reduction}
This paper presents new ALMA Band 7 observations taken as part of the program 2023.1.00428.S (PI A. Miotello). For both science goals, the setup configuration consists of five spectral windows (SPWs): two centered on the $^{12}$CO $(J=3-2, \nu_{\rm rest}=345.795\,{\rm GHz})$ and $^{13}$CO $(J=3-2, \nu_{\rm rest}=330.587\,{\rm GHz})$, one targeting CS $(J=7-6, \nu_{\rm rest}=342.882\,{\rm GHz})$, one targeting HC$^{15}$N $(J=4-3, \nu_{\rm rest}=344.200\,{\rm GHz})$, and an additional wideband spectral window for the continuum $(\nu_{\rm center}=332.900\,{\rm GHz})$. All windows are configured in Frequency Division Mode (FDM). The details of the spectral setups used in the observations are summarized in Tab.~\ref{table:setup}.

The targets were observed in two scheduling blocks, corresponding to the two science goals of the project. The first one was executed once on June 10 2024, while the second was observed four times between August 1 and 14 2024. The observations were carried out with the 12~m array, using up to 45 antennas in Band 7 at an angular resolution of $0\farcs25$. Each successful EB lasted from about 7 to 45 minutes on source. A summary of the observational setup and conditions is provided in Tab.~\ref{table:conditions}.

The data were pipeline-calibrated using the Common Astronomy Software Applications package (CASA, version 6.5.4-9, \citealt{CASATeam22}). The measurement sets were initially delivered as concatenated datasets including all execution blocks. These were then split by execution block and by source using the \texttt{split2} task, creating individual datasets for each target. Following the DSHARP alignment strategy \citep{Andrews18}, the phase center of each EB was shifted to the continuum peak using the \texttt{phaseshift} task, and then aligned relative to the brightest EB to ensure consistent astrometry across all datasets. To place all executions in a consistent celestial frame, the \texttt{fixplanets} task was used to realign the coordinate systems based on the updated phase centers. This process ensured that all datasets were centered consistently. After applying all corrections, the datasets corresponding to each source were re-concatenated to produce a final, aligned dataset ready for imaging. 

Image reconstruction was performed using \texttt{tclean}. The images were deconvolved interactively to carefully control the CLEANing process, considering the very faint emission channel by channel. The velocity widths were set to $0.25\,{\rm km}\,{\rm s}^{-1}$ for $^{12}$CO and $0.50\,{\rm km}\,{\rm s}^{-1}$ for $^{13}$CO, which are larger than the native spectral resolution of the data. This choice was motivated by the very low signal-to-noise ratio per channel at the native resolution, which prevented reliable identification of the line emission in the image cubes. The imaging was performed using a multiscale deconvolver, a cell size of approximately 1/8 of the beam semi-minor axis. CS and HCN  will be analyzed in a forthcoming paper. 

Lastly, moment zero maps were generated using the \texttt{bettermoments} package \citep{Bettermoments}, collapsing the cubes over a velocity range chosen to include only the channels with emission, plus two channels on either side. The $^{12}$CO and $^{13}$CO spectra were extracted from the image cubes by defining a two-dimensional elliptical mask centered on the line emission, with semi-axes and orientation chosen to match the inclination and position angle derived from the continuum visibilities and to fully enclose the detected CO emission. The final continuum and CO integrated intensity maps and integrated spectra are shown in Fig.~\ref{figure:sample}. 

\begin{figure*}[htb!]
    \centering
    \includegraphics[width=1.5\columnwidth]{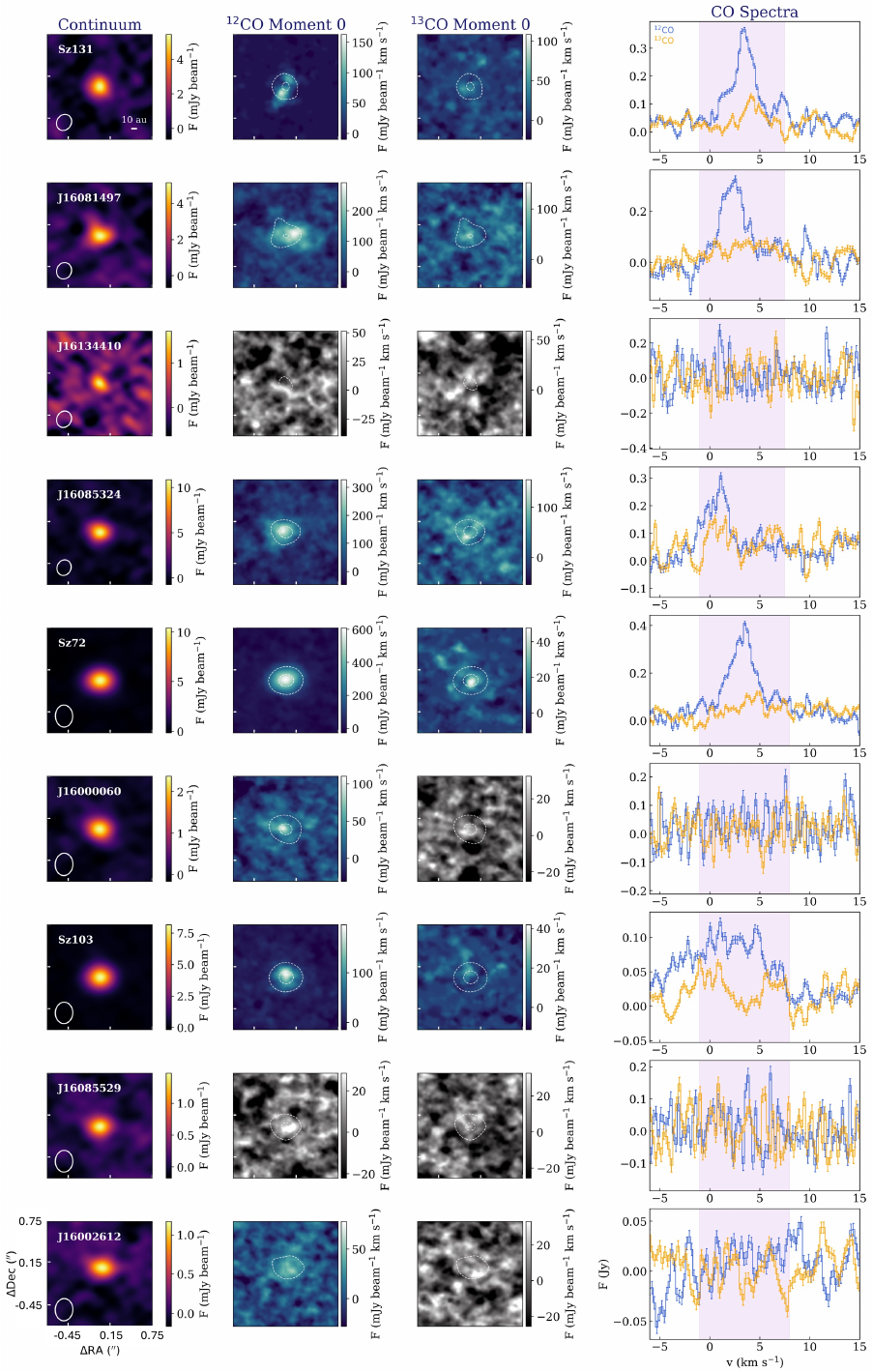}
    \caption{ALMA Band 7 observations of the 17 Lupus disks in our sample. From left to right: continuum, $^{12}$CO, and $^{13}$CO moment 0 maps, and the integrated spectra for both the molecules. In CO maps, the white contours indicate the continuum emission at the 3 and $10\sigma$ levels. Gray panels indicate non detections. In spectra, the shaded region represents the velocity ranges used to create the moment 0 maps. The synthesized beam is shown as a white ellipse in the lower right corner of the continuum images.}
    \label{figure:sample}
\end{figure*}

\begin{figure*}[htb!]
    \centering
    \addtocounter{figure}{-1}
    \includegraphics[width=1.5\columnwidth]{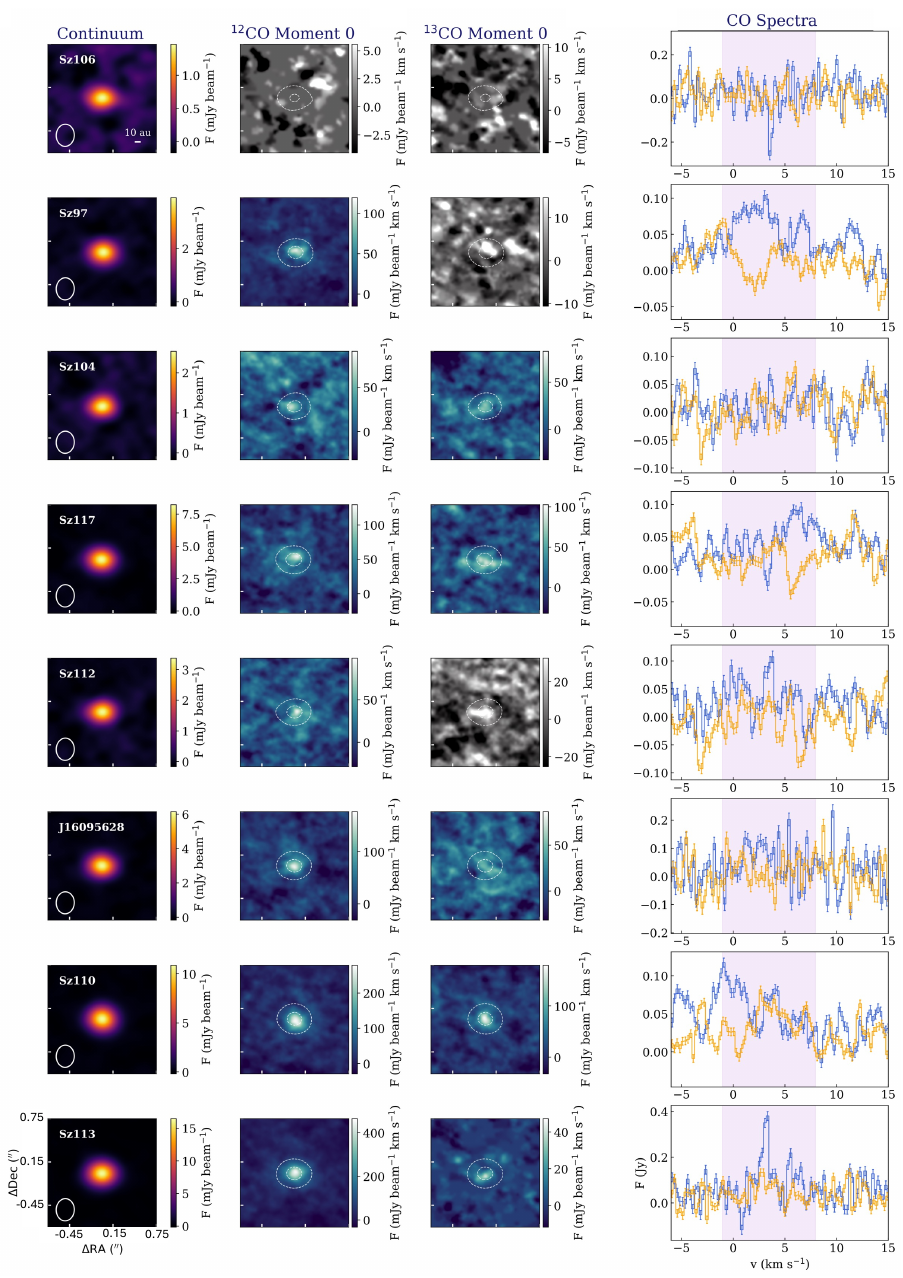}
    \caption{Figure~\ref{figure:sample} (continued).}
    \label{figure:sample1}
\end{figure*}

\section{Methods} \label{section3}
In this section, we introduce the methods adopted to measure the CO fluxes and the CO and dust disk sizes in our sample.

\subsection{CO fluxes estimate: the GoFish tool}
The disks in our sample are faint in CO emission. In such a low signal-to-noise ratio (S/N) regime, techniques that enhance line detectability are essential.
For our analysis, we used the Python-based package GoFish \citep{GoFish}. This tool exploits the predictable, ordered velocity field of disks in Keplerian rotation to align and coherently stack spectra extracted at different positions across the disk. The spectra are then azimuthally and/or radially stacked, effectively compressing the emission into a narrower velocity range. This increases the peak intensity of the stacked spectrum without altering the total integrated flux, resulting in a significant gain in S/N ratio. The alignment process requires knowledge of the disk inclination, position angle and central stellar mass and distance, which define the projected Keplerian velocity field. In the case of planet-forming disks this technique was pioneered by \cite{Yen16}, \cite{Matra17} and \cite{Teague21}, and has since been widely adopted to enhance faint line emission \citep[e.g.,][]{Rampinelli24, Zagaria2025, Rodriguez25, Deng25}. It is particularly well suited to our sample, where the CO emission is expected to be compact and faint.

\begin{figure}[htb!]
\centering
\subfigure{
   \includegraphics[width=8cm]{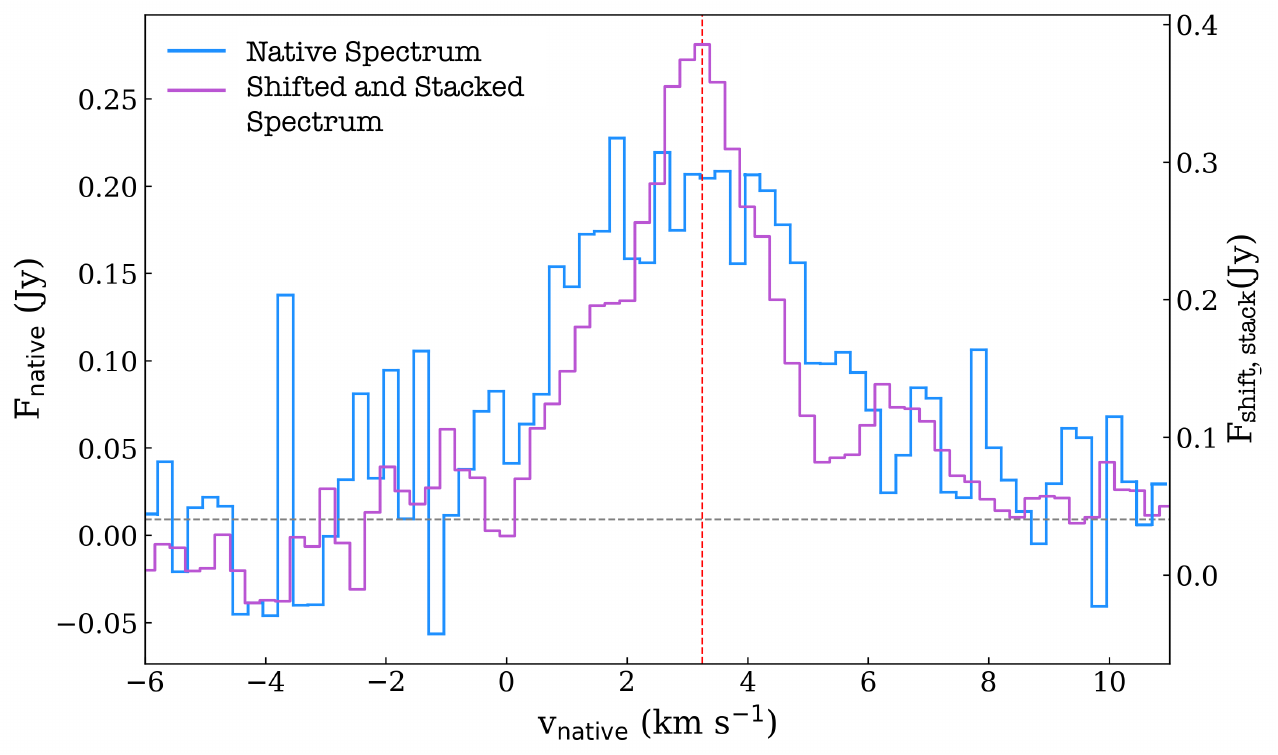}
}
\subfigure{
   \includegraphics[width=8cm]{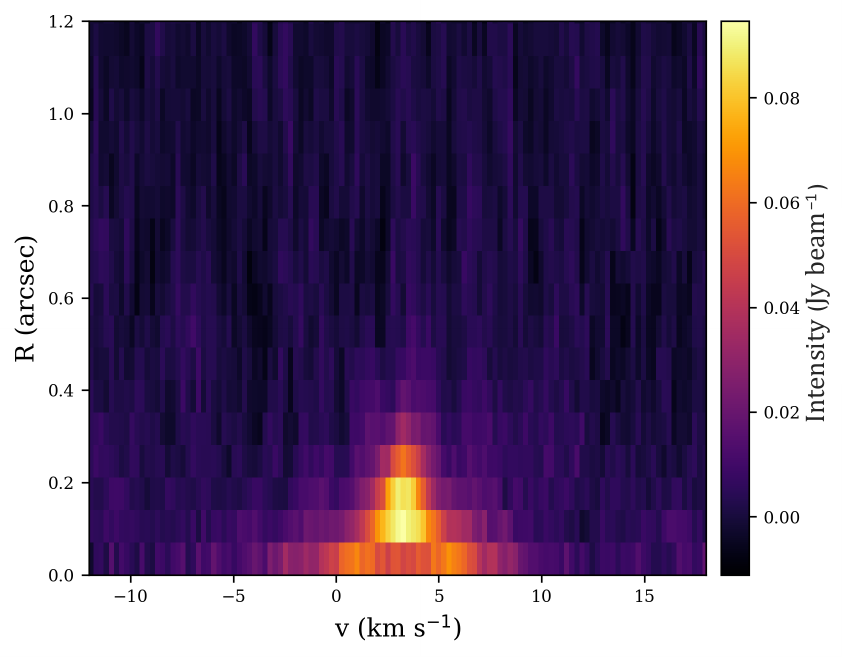}
}
\caption{\textit{Top panel:} Native (blue) and shifted-and-stacked (purple) integrated spectra for Sz~72. The vertical dashed red line and the horizontal dashed gray line mark the systemic velocity and the RMS noise level of the source, respectively. \textit{Bottom panel:} Teardrop plot for the $^{12}$CO emission in Sz~72, displaying the spectra as a function of radius after correcting for the expected Keplerian rotation.} 
\label{figure:gofish}
\end{figure}

Position angles (PA) and inclinations ($i$) were estimated from continuum visibilities fits of our continuum visibilities (more details in Section~\ref{dust_fit}). For stellar masses and distances, instead, we adopted the values from \citet[][Table 1]{Manara23}. For the analysis, we used the \texttt{integrated$\_$spectrum} function to extract the CO line fluxes. An example of the resulting integrated spectrum is shown in the top panel of Fig.~\ref{figure:gofish}. To verify the success of the alignment and stacking procedure, we used the \texttt{plot$\_$teardrop} function, which produces a radially resolved position–velocity diagram. When the alignment is correct, the emission appears symmetric and centered around the systemic velocity, as shown in the bottom panel of Fig.~\ref{figure:gofish}. We measured the CO flux by integrating the shifted-and-stacked velocity spectrum where the emission was clearly detectable above the local noise. We identified these channels by comparing the spectrum with the RMS per channel and marking those significantly ($\geq3\sigma$) above the noise level as signal. We estimated then the uncertainty on the integrated flux empirically: we shifted the same spatial mask used to extract the spectrum over several emission-free regions of the moment 0 map, extracted the enclosed flux, and considered the standard deviation of these values as the uncertainty in the integrated flux. The final values are listed under the heading ``Shift and stacked line emission'', in Tab.~\ref{table:sample}.

While the GoFish tool provided reliable CO fluxes for eight sources in our sample--the brightest ones that were (at least marginally) spatially resolved in the gas--it could not be successfully applied to the remaining targets. Indeed, if the disk is not spatially resolved, the shift-and-stack technique becomes ineffective. This is because the velocity shifts applied are incorrect for unresolved emission, causing the signals to de-cohere, ultimately lowering the peak S/N of the resulting spectrum rather than enhancing it. In these cases, either the disk geometry was poorly constrained (e.g., Sz~110), or the CO emission was too faint or was entirely undetected, or both. The procedure adopted for the latter sources is explained in the following section.

\subsection{CO fluxes estimate: image plane analysis}
For those unresolved sources, where we were not able to shift and stack the CO lines, we extracted integrated line fluxes directly from the integrated intensity maps using elliptical masks that matched the size and position angle of the synthesized beam and were sufficient to enclose the full emission. The uncertainty of these fluxes was estimated empirically by applying the same mask to multiple emission-free regions of the integrated intensity image, while avoiding the edges. The standard deviation of the extracted fluxes in each region was then adopted as the flux uncertainty. The resulting values are listed under the heading flux extraction from the integrated intensity maps, in Tab.~\ref{table:sample}. 

\subsection{Disk Gas Radii Estimate} \label{gas_fit}
We estimated the gas disk sizes following \cite{Sanchis21} by fitting a two dimensional elliptical Gaussian to the CO integrated intensity maps using the CASA task \texttt{imfit}, which provided an estimate of a deconvolved Gaussian Full Width at Half Maximum (FWHM). In the literature, disk sizes are commonly defined as the radius enclosing a fixed fraction of the total flux, typically the 68$\%$ ($R_{68}$) and the 90$\%$ ($R_{90}$) for protoplanetary disks \citep[e.g.,][]{Tripathi17, Hendler20}. We therefore converted the measured FWHM into these radii using the standard relations for a Gaussian profile (Eq. 1 and 2 of \citealt{Sanchis21}). The uncertainties on the radii were propagated from the errors on the FWHM of the major axes provided by \texttt{imfit}. We preferred estimating disk radii using \texttt{imfit}, rather than a curve of growth method because this approach is less sensitive to noise, particularly for $R_{90}$.

\subsection{Disk Dust Radii Estimate} \label{dust_fit}

To estimate the disk dust radius, we modeled the millimeter continuum emission using \texttt{GALARIO} \citep{Galario} to fit the observed continuum visibilities. \texttt{GALARIO} efficiently performs forward modeling of interferometric data by computing synthetic visibilities from a parametric brightness distribution and comparing them directly with the observations.

For each source, we extracted the continuum visibilities after carefully flagging the frequency ranges where we detect or expect emission from the targeted CO isotopologues, CS, and HCN lines. We described the radial distribution of the continuum emission with a Gaussian profile. We constrained the model parameters using a Bayesian inference framework implemented with the \texttt{emcee} Markov Chain Monte Carlo (MCMC) sampler \citep{Foreman13, Foreman19}. The free parameters included the flux normalization, Gaussian width, disk inclination, position angle, and positional offsets in right ascension and declination. We chose uniform priors within physically motivated ranges and sampled the flux normalization in logarithmic space to account for the wide dynamic range in flux densities.

We generated synthetic visibilities for each parameter set and compared them directly with the data through a $\chi^{2}$ likelihood function. We explored the posterior distributions using 120 walkers and stopped after 5000 steps. This yielded robust estimates of the best-fit parameters and their associated uncertainties. We then derived the disk dust size, defined as the disk radius enclosing 68\% and 90\% of the total fitted flux.

\section{Results} \label{section4}
The key question we aim to address is whether the faint CO emission from the disks in our sample can be explained simply by their small spatial extent. If both $^{12}$CO and $^{13}$CO emission are optically thick, their line luminosities are expected to scale primarily with the emitting surface area. In that case, variations in disk size would produce proportional changes in both isotopologues, leading to an approximately linear correlation between their line luminosities. In this section, we present the measured disk sizes and compare the radial extents of the gas and dust emission.

\subsection{Integrated Fluxes and Disk Extents}
We detected $^{12}$CO emission (above the $3\sigma$ level) in all but four sources, while $^{13}$CO emission, as expected, is always weaker than $^{12}$CO, resulting in seven non-detections. In agreement with the previous non-detections from \cite{Ansdell16, Ansdell18}, the CO emission is faint ($L_{\rm ^{12}CO}\leq10^9\,{\rm mJy}\,{\rm km}\,{\rm s}^{-1}\,{\rm pc}^2 \simeq 8.5 \times 10^{-2} L_{\odot}$). The final values are listed in Tab.~\ref{table:sample}. 

\begin{figure*}
    \centering
    \includegraphics[width=2\columnwidth]{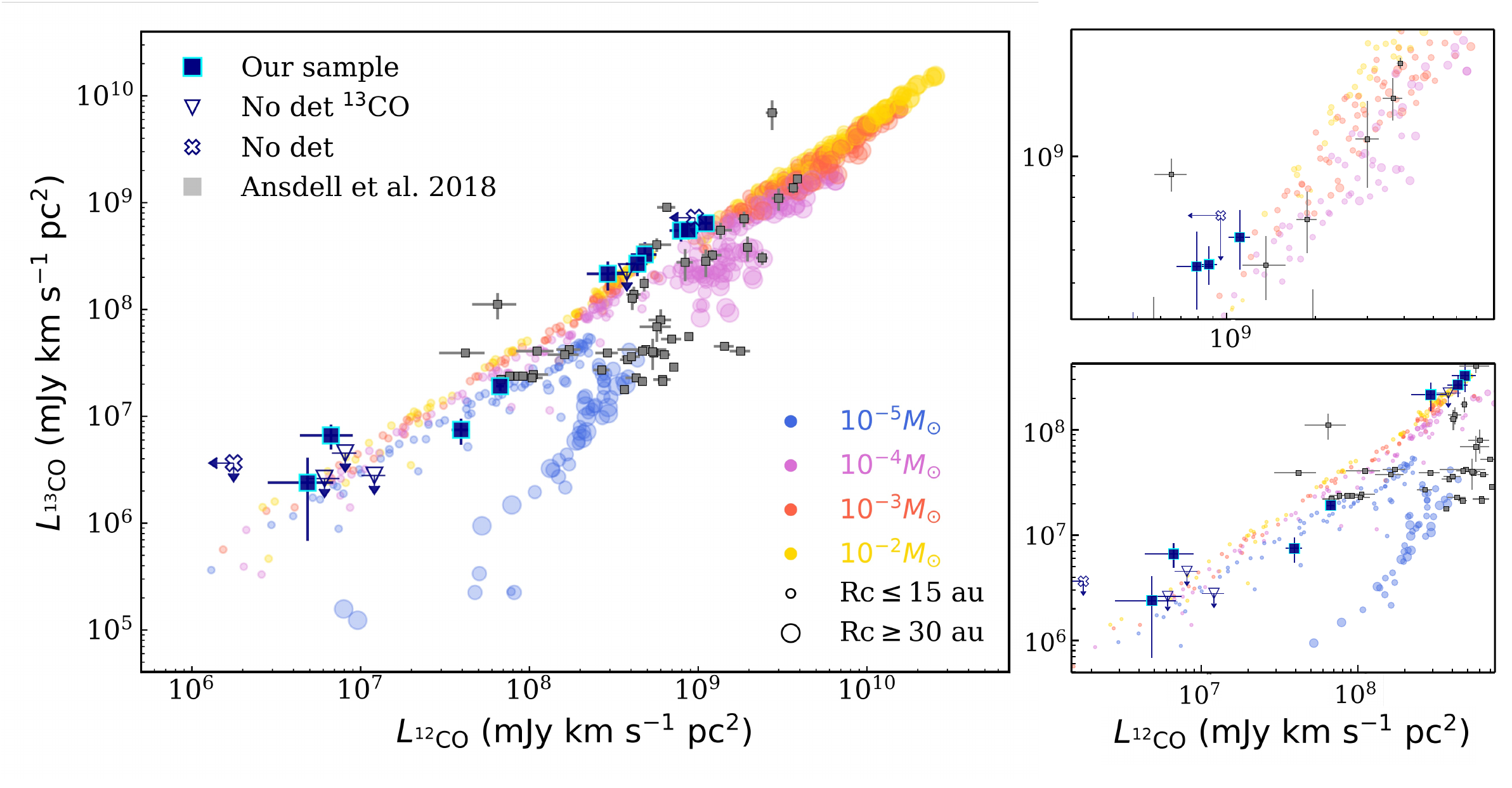}
    \caption{Comparison of Lupus disk CO line luminosities with physical–chemical model predictions. The DALI models from \citet{Miotello21} are shown as background dots; different colors indicate different disk gas masses, and the symbol size scales with disk radius. The Lupus sources from \citet{Ansdell18} are overplotted as gray squares. For our sample, blue sqares mark detections, open triangles indicate non-detections in $^{13}$CO, and open crosses represent non-detections in both CO isotopologues. The two panels on the right show zoomed-in views: the top one show the targets in the first SG, the bottom one the targets of the 2nd SG.}
    \label{figure:luminosities_plot}
\end{figure*}

\cite{Miotello16, Miotello21} created two grids of physical-chemical disk models run with DALI (Dust And LInes, \citealt{Bruderer12}), the first one with more extended disks models (scale radius $\geq30\,{\rm au}$), and the second including much more compact disk models (scale radius $\leq15\,{\rm au}$). Both model grids span the same range of disk masses ($10^{-5} M_\odot \leq M_{\rm disk} \leq 10^{-4} M_\odot$). Such model grid results were compared to one another in \citet{Miotello21}, where a bimodal distribution was seen in the $^{12}$CO vs. $^{13}$CO integrated luminosity plane, as shown also in Fig.~\ref{figure:luminosities_plot} (where models are color-coded by total disk mass and size-coded by their scale radius). In the case of compact disks, the CO line luminosities are linearly correlated across more than four orders of magnitude. This is because both $^{12}$CO and $^{13}$CO are optically thick at any disk mass if the disks are so compact, and the CO fluxes scale with the emitting area. Therefore, lower luminosities correspond to smaller disks in size. On the other hand, for extended disks, at lower disk masses for fainter (but still optically thick) $^{12}$CO emission, the $^{13}$CO emission becomes progressively optically thinner. The correlation steepens and, in this case, faint emission is due to an extended gaseous structure with low surface density. \cite{Miotello21} could not discriminate between the two scenarios, as observations in Lupus were not deep enough to probe the region in the $^{12}$CO vs. $^{13}$CO integrated luminosity plane where the bifurcation between these models happens, which we now can probe with the new data presented in this paper.

In the same diagram, we overplotted the CO fluxes of our 17 targets, as well as the previously observed disk population from \citet{Ansdell18}, as gray markers. The non-detections in the latter disk sample are consistent with both the extended and the compact scenarios \citep[see][]{Miotello21}. Our survey is about one order of magnitude deeper and shows that all the detections are consistent with the $^{12}$CO and $^{13}$CO isotopologues being optically thick, and a few of them are consistent with compact disk models (in the bottom left region of Fig.~\ref{figure:luminosities_plot}), suggesting that some of these disks may be faint because they are compact. Our only three non-detections remain compatible with both scenarios.

Nevertheless, we aim to measure the gas disk radii, where possible, to further test this scenario, while acknowledging that some disks could still be more extended, with a faint optically thin halo that remains below our detection limits and contributes little to the total flux. We derived both the gas and dust radii for the objects in our sample, following the procedures described in Subsections~\ref{gas_fit} and \ref{dust_fit}. The resulting $R_{\rm CO,68}$ and $R_{\rm CO,90}$ are listed in Tab.~\ref{table:sample}. For the gas specifically, we note that the Gaussian fitting procedure could not be performed for J16000060$-$4221567 and Sz~104, because the CO emission appears unresolved, preventing \texttt{imfit} from returning meaningful unconvolved sizes. For these disks, Tab.~\ref{table:sample} shows the upper limits defined as half of the minor axis of the beam. We also attempted to derive gas radii from the $^{13}$CO emission using the same fitting procedure. However, due to the lower signal-to-noise ratio and the compact nature of the emission, none of the disks are spatially resolved in this tracer.

For the resolved sources, we estimated a median gas disk size $\langle R_{\rm CO,68}\rangle=23.62 \pm 2.65\,{\rm au}$ and $\langle R_{\rm dust,68}\rangle=12.75 \pm 0.58\,{\rm au}$. Such values are significantly smaller than the typical gas radii found for Lupus disks thus far (e.g., median value of $\sim$120~au for \citealt{Sanchis21}, and $\sim$78~au for \citealt{Trapman25_sizes}). Our measurements confirm that the disks in our sample are radially compact. This finding strengthens the possibility that the low CO luminosity observed could be the result of a limited surface area over which CO remains optically thick.

\subsection{Gas-to-dust Size Ratio}
The gas-to-dust size ratio $R_{\rm CO}/R_{\rm dust}$ is an important diagnostic for dust processing. In models that include radial drift and grain growth, millimeter-sized dust particles migrate inward relative to the gas, leading to a natural decoupling between the spatial distributions of gas and solids. As a first approximation, the gas radius traced by optically thick CO emission is set by the location where the CO column density reaches the photodissociation threshold, and is therefore only weakly affected by dust evolution. In contrast, the dust outer radius is directly shaped by the radial redistribution of millimeter-sized grains. As a consequence, radial drift leads to systematically smaller dust radii compared to gas radii, producing larger values of $R_{\rm CO}/R_{\rm dust}$ than in disks where dust remains well coupled to the gas. Numerical models \citep{Trapman19} show that including dust evolution reduces the observed millimeter dust radius, while the CO emission remains relatively extended, thereby increasing the size ratio compared to models without drift (where only optical depth effects are considered). Under this framework, one may expect a correlation between the size ratio and disk radii if radial drift is the dominant mechanism shaping disk structure: disks with more efficient drift should exhibit smaller dust radii at fixed gas radii, resulting in an increasing $R_{\rm CO}/R_{\rm dust}$ as the dust radius decreases (or as the gas radius becomes larger relative to the dust). This expectation motivated earlier searches for size–radius correlations in both models and observations \citep[e.g.,][]{Sanchis21,Toci21}. However, both models and observations indicate that for the majority of disks the size ratio lies within a relatively narrow range of $\sim 2$–$4$, which can often be reproduced without invoking strong radial drift once optical depth effects are included. In fact, \citet{Trapman19} models suggested that ratios $\geq 4$ provide unambiguous evidence for significant dust evolution, while more moderate values may also arise from optical depth differences between CO and dust continuum. So far, the results for the most luminous disks in Lupus have shown that disks are systematically larger in gas than in dust, with a median ratio of $R_{\rm CO}/R_{\rm dust} \approx 2.5-3$ \citep{Sanchis21}. More recently, \citet{Pinilla25} found a similar median size ratio of $2.7^{+0.3}_{-0.4}$ in Upper Scorpius. This trend appears to be robust across different star-forming environments: analyzing a sample of protoplanetary disks in several nearby SFRs, \citet{Long22} similarly concluded that CO emission is universally more extended than the continuum emission by an average factor of $2.9 \pm 1.2$.

\begin{figure}
    \centering
    \includegraphics[width=\columnwidth]{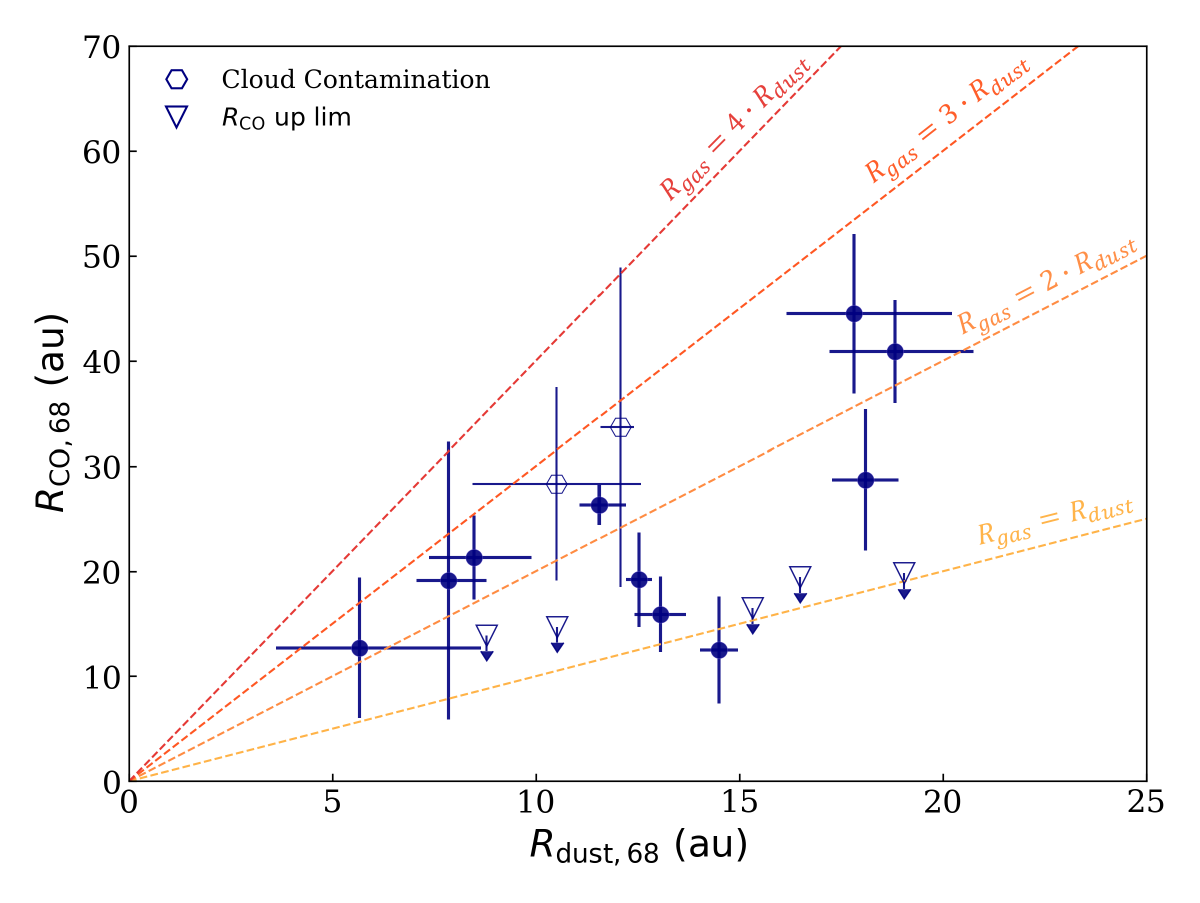}
    \caption{Comparison between dust and gas sizes in our sample. Lines represent size ratios from 1 to 4.}
    \label{figure:size_ratio}
\end{figure}

In our analysis, we compute the size ratio using $R_{\rm CO, 68}$ and $R_{\rm dust, 68}$, to be consistent with \cite{Sanchis21}. The size ratios of our faint structures are listed in Tab.~\ref{table:sample} and shown in Fig.~\ref{figure:size_ratio}. Detections are shown as blue circles, while upper limits on $R_{\rm CO}$ are displayed as triangles. The two sources Sz~112 and J16002612–4153553, marked with hexagons, are affected by cloud contamination. Consequently, their measured CO brightness and radius are likely underestimated and should be interpreted as lower limits in any subsequent analysis. Most of our disks exhibit size ratios between 2 and 3, in good agreement with the median value reported by \citet{Sanchis21}, and independent of the fraction of flux used to estimate the radii. In fact, considering $R_{\rm CO,90}$ and $R_{\rm dust,90}$ does not alter our results. Notably, none of our sources reach size ratios above 4, which are generally considered a strong indication of radial drift \citep{Trapman19, Toci21}, supporting the conclusion that these disks are not drift-dominated.

\begin{figure*}[htb!]
    \centering
    \includegraphics[width=2\columnwidth]{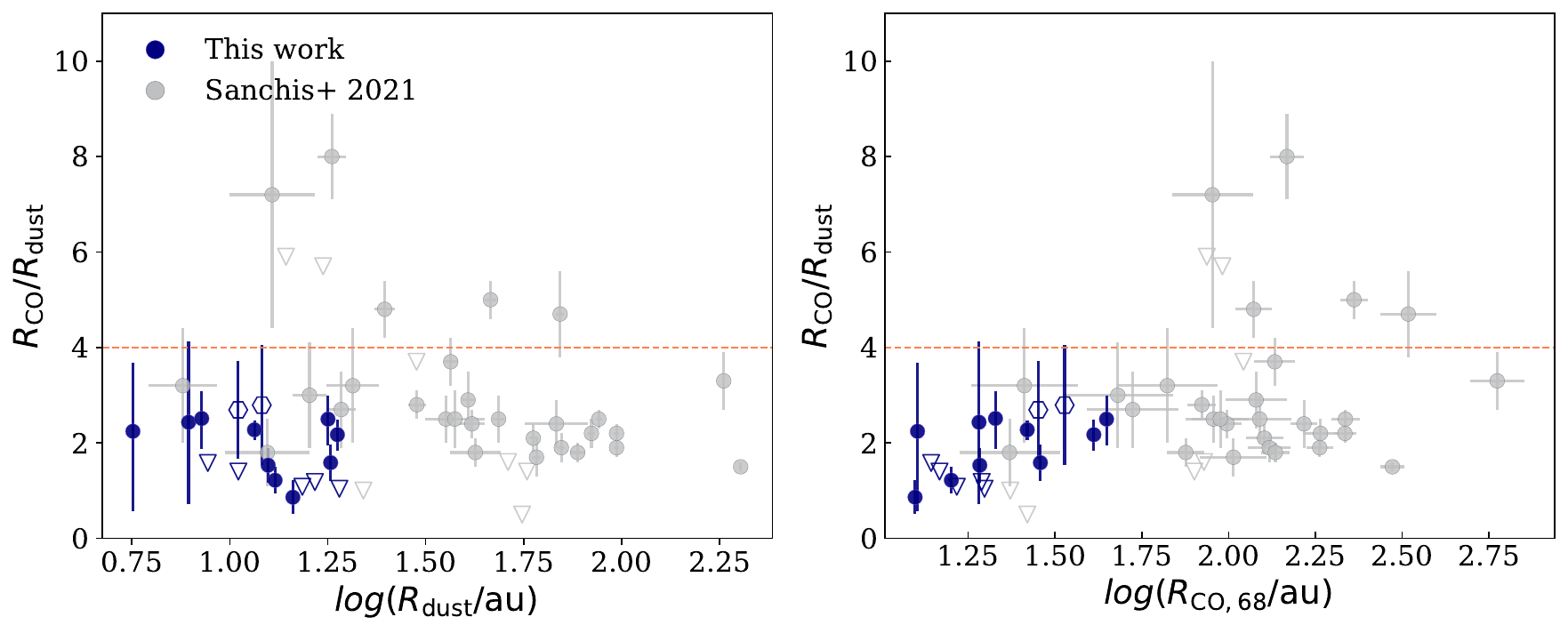}
    \caption{Size ratio as a function of the dust and the $^{12}$CO radii enclosing the 68$\%$ of the total flux. The results of this work are plotted as blue dots, those by \cite{Sanchis21} as gray dots. The upper limits are indicated as triangles.}
    \label{figure:ratio_vs_radii}
\end{figure*}

\begin{figure}
    \centering
    \includegraphics[width=\columnwidth]{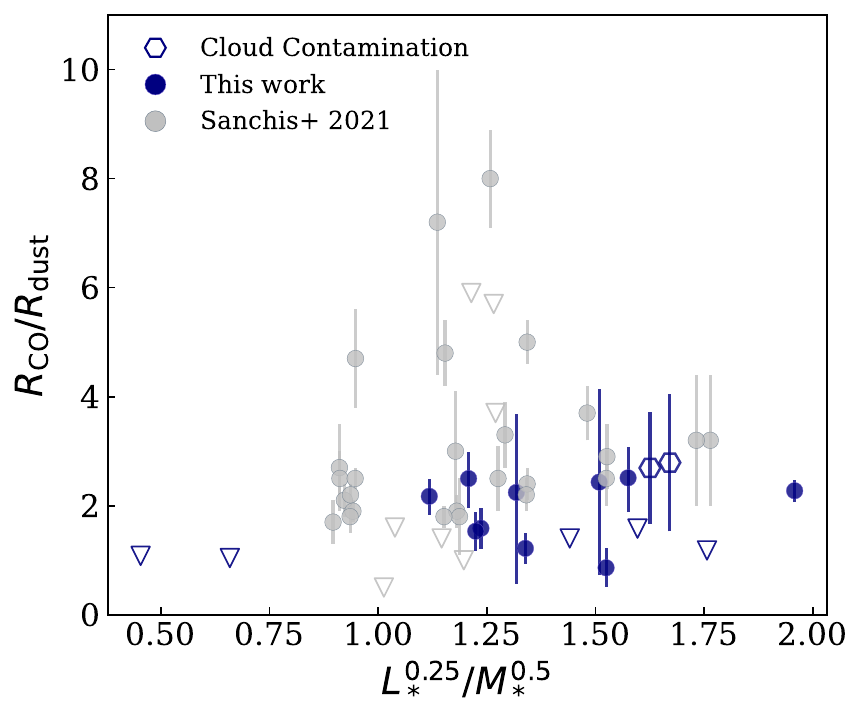}
    \caption{Gas-to-dust size ratio ($R_{\rm CO}/R_{\rm dust}$) as function of $L_{\star}^{1/4}/M_{\star}^{1/2}$, a proxy of the dust radial drift. Blue and gray dots represent our sample and that of \citet{Sanchis21}, respectively. The upper limits are indicated as triangles.}
    \label{figure:ratio_stellar}
\end{figure}

In Fig.~\ref{figure:ratio_vs_radii} we plot the size ratio as a function of the dust and $^{12}$CO disk size, and in Fig.~\ref{figure:ratio_stellar} as a function of the expected dust drift velocity. In both figures we combined our sample (in blue) and the \citet{Sanchis21} one (in gray) to probe a broad range of stellar and disk properties. We first investigated the presence of a correlation between the size ratio and the logarithm of the dust and gas sizes using the Bayesian linear regression method implemented in \texttt{linmix}\footnote{\url{https://github.com/jmeyers314/linmix}} \citep{Kelly_2007}. This analysis yielded a correlation significance close to zero in both cases, indicating no statistically significant trend. From a theoretical perspective, if radial drift were the dominant mechanism shaping dust evolution, one would expect a linear correlation between the size ratio and the dust and gas sizes: more efficient drift produces smaller dust radii while the gas radius remains almost unaffected, leading to higher $R_{\rm CO}/R_{\rm dust}$ values. The absence of a statistically significant correlation in Fig.~\ref{figure:ratio_vs_radii} therefore suggests that radial drift is not efficient across our sample.

We then explored possible connections between the size ratio and the expected dust drift velocity, $v_{\rm drift}$. \citet{Pinilla13,Pinilla22} suggested that $v_{\rm drift}$ scales with stellar mass and luminosity as $L_\star^{1/4} / M_\star^{1/2}$. If radial drift were the dominant mechanism shaping dust evolution, a positive correlation between $R_{\rm CO}/R_{\rm dust}$ and $v_{\rm drift}$ would therefore be expected. However, no such trend can be seen in Fig.~\ref{figure:ratio_stellar}, strengthening our conclusion that most Lupus disks are not drift-dominated. We note, however, that our sample spans a wide range of stellar masses and luminosities, and additional factors (e.g., the gas surface density profile or the dust-to-gas ratio) could in principle weaken any intrinsic correlation.Alternatively, the compact disks may host unresolved substructures that trap dust and limit its inward migration, even as $L_{\star} / M_{\star}$ increases \citep{Toci21}.

\begin{figure}[htb!]
    \centering
    \includegraphics[width=\columnwidth]{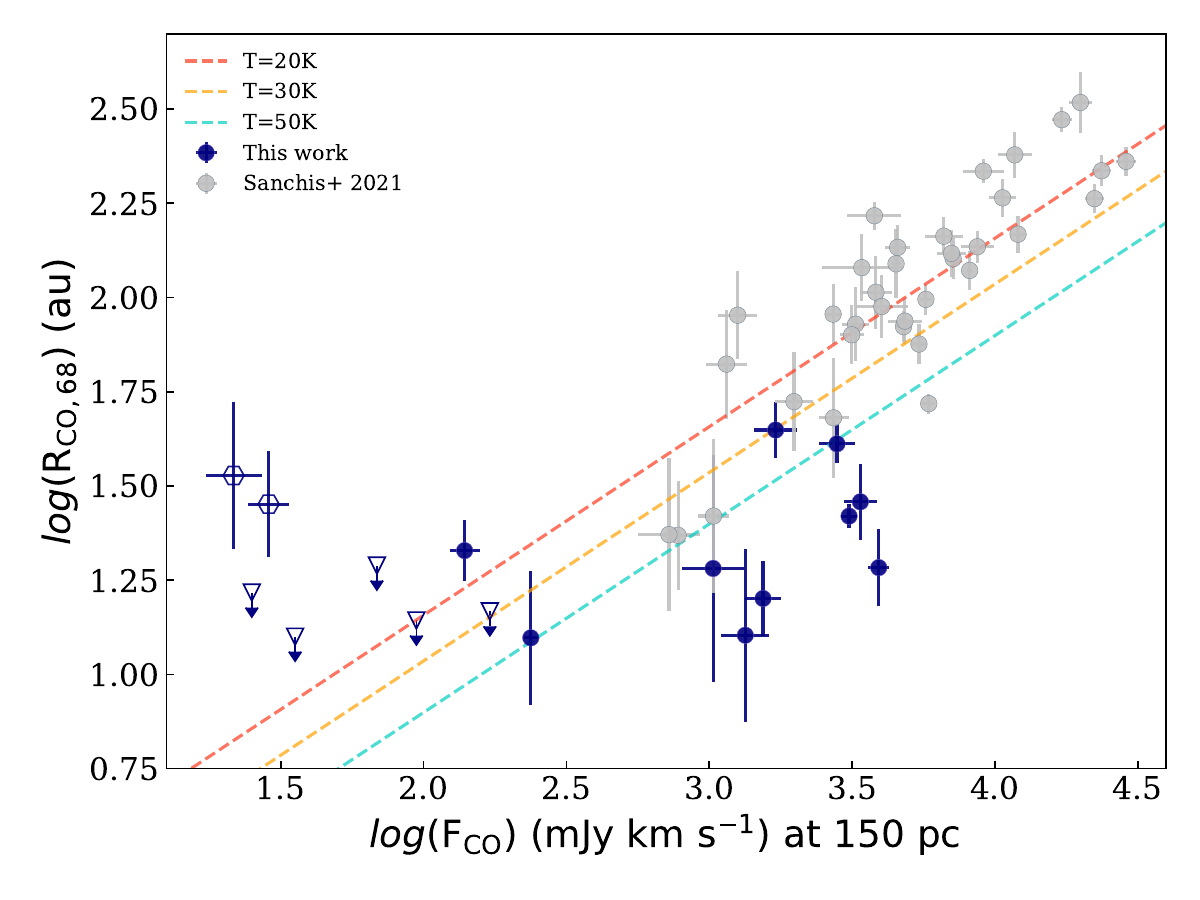}
    \caption{Relation between $R_{68, \rm CO}$ and the $^{12}$CO line luminosity, scaled at the median distance of $150\,{\rm pc}$ and corrected for inclination. Our sample is shown as blue dots, while the sample of \cite{Sanchis21} as gray ones. The overplotted dashed lines are optically thick CO models with an average temperatures of 20 K (coral), 30 K (yellow), and 50 K (green).}
    \label{figure:lum_vs_rad}
\end{figure}

Lastly, Fig.~\ref{figure:lum_vs_rad} shows $R_{\rm CO,68}$ against $F_{\rm CO}$ rescaled to a common distance of $150\,{\rm pc}$ for our sample (in blue) and the \citet{Sanchis21} one (in gray), and corrected for inclination by dividing $F_{\rm CO}$ by $\cos i$ (in radians). Since \citet{Sanchis21} analyzed CO ($J=2-1$) observations, we rescaled their CO fluxes by $(\nu_{J=3-2}/\nu_{J=2-1})^2\approx (345.796/230.538)^2\approx2.25$, corresponding to the ratio of the fluxes in the two transitions under the assumption of optically thick emission. Overplotted are reference curves for optically thick emission at different temperatures. To construct these dashed curves, we created a grid of optically thick, isothermal emission profiles, $I_{\rm CO}(R) = B_{\rm \nu}(T_{\rm CO})$, where $\nu$ is the frequency of the $^{12}$CO $(J=3-2)$ transition and $T_{\rm CO}$ is equal to $20,\, 30,\,{\rm and}\,50\,{\rm K}$, integrated over a $\Delta v=0.25\,{\rm km/s}$.Overall, we observe a positive correlation between luminosity and size, which is consistent with the findings of \cite{Sanchis21} and \cite{Long22}. Most of our disks lie very close to this relationship. However, a subset of objects appears significantly more luminous than expected from their measured sizes. This excess emission could be explained by higher disk temperatures, as suggested by our reference curves, or by the possibility that these disks are more extended because of faint, low-surface-brightness emission being missed in our data.

\section{Discussion} \label{section5}
Our observations show that some of the faintest disks in Lupus are consistent with being optically thick in $^{12}$CO and $^{13}$CO, and radially compact in both gas and dust, with $R_{\rm CO, 68} \leq 40\,{\rm au}$ and $R_{\rm dust, 68} \leq 20\,{\rm au}$. Our results support the interpretation that their faint CO emission is predominantly due to their compactness and not necessarily to a reduced surface brightness by extreme carbon depletion. However, some levels of depletion cannot be ruled out. Hereafter we discuss the broader implications of our results on disk evolution. We stress that our goal is not to explain why the faint Lupus discs analyzed in this paper are compact, but rather what constraints (e.g., on initial conditions and evolution mechanisms) their small sizes provide.

\subsection{Disk evolution}
Protoplanetary disk evolution and dispersal are among the most debated topics in planet formation. Over the years, two main frameworks have been proposed. In the traditional viscous scenario \citep{Shakura&Sunyaev73,LyndenBell74,Hartmann98,Lodato17,Rosotti17}, evolution takes place by turbulent angular momentum redistribution while thermal winds regulate disk dispersal. More recently, an alternative scenario was revived in which evolution and dispersal are both regulated by angular momentum removal driven by magneto-thermal winds \citep{Blandford&Payne82,Ferreira97,Bai&Stone13,Suzuki16,Tabone22a}. One of the key most diverging predictions of these models is that gas disk sizes should increase with time in the viscous case \citep[cf., e.g.,][]{Trapman20}, but stay constant or shrink in the MHD-wind case \citep[e.g.,][]{Trapman22}. Therefore, comparing predicted and observed disk sizes could provide key information of what mechanisms are governing disk evolution. Since this comparison is most significant at a population level, it naturally requires some degree of simplification in our models, which nonetheless comes with the reward of identifying key ballpark demographic trends essential to interpret population observations.

Our Lupus sample is particularly useful in this regard for two main reasons: (1) in this region, disk evolution is likely not substantially affected their surrounding environment (in the sense that our disks are intrinsically small and not truncated by e.g., external photoevaporation or dynamical encounters; cf., subsection~\ref{section2.1}), suggesting that 
our sources are compact primarily as a consequence of internal (viscous- or MHD-driven) evolution, and (2) such compact disks allow us to test model predictions in a challenging region of the parameter space, rarely probed to date.

\begin{figure*}
    \centering
    \includegraphics[width=15cm]{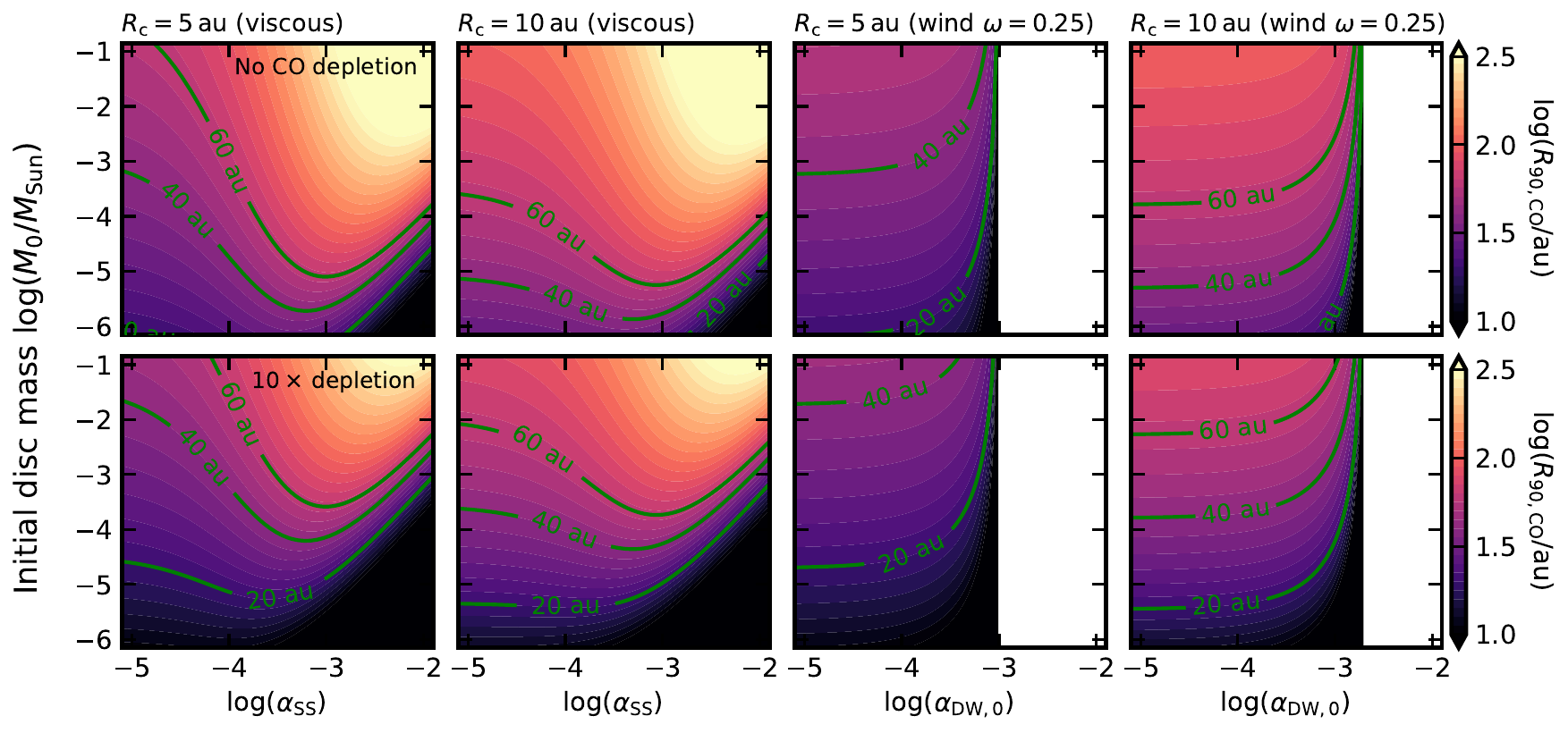}
    \caption{Toy-model-based gas disk sizes ($R_{\rm CO,90}$) as a function of the initial disk mass ($M_0$) and evolution efficiency ($\alpha$) for different initial disk sizes ($R_{\rm c}=5$ and $10\,{\rm au}$, odd and even columns) CO depletion factors (0 and 10, top and bottom rows) in the viscous and MHD-wind case. The green curves display contours compatible with the gas size range spanned by our data (cf., Tab.~\ref{table:sample}).}
    \label{figure:models}
\end{figure*}

We used the analytical prescription of \citet[][see also \citealt{Toci23}]{Trapman23} to estimate the gas disk size ($R_{\rm 90,CO}$) of a set of models, whose initial disk size ($R_{\rm c}$) and masses ($M_0$) were evolved in the viscous (cf., \citealt{Lodato17}) and MHD-wind case (cf., \citealt{Tabone22a}, in the finite-time dispersal case with $\omega=0.25$, \citealt{Tabone25}) with different efficiency ($\alpha_{\rm SS}$ and $\alpha_{\rm DW,0}$) up to $2\,{\rm Myr}$ (the median age of the region, \citealt{Alcala14,Luhman20}). We explored a wide range of initial disk masses and $\alpha$ parameters for a few representative initial disk sizes. Our results are displayed in the top row of Fig.~\ref{figure:models}.

In the viscous case (leftmost panels), $R_{\rm 90,CO}$ increases with $\alpha_{\rm SS}$ because more turbulent disks expands more. However, for low disk masses or a very high $\alpha_{\rm SS}$ the trend changes. In this regime, our models have expanded so much that $R_{\rm 90,CO}$ falls in the power-law region of the density profile (rather than its exponential tail) and thus decreases when the disk viscously expands (as was already noticed by \citealt{Trapman20,Toci21,Toci23}). Instead, in the MHD-wind case (rightmost panels) $R_{\rm 90,CO}$ stays constant with $\alpha_{\rm DW,0}$ but drastically decreases for $\alpha_{\rm DW,0}\approx10^{-3}$ as a consequence of the abrupt disk dispersal (cf., \citealt{Tabone22a,Trapman22}. For larger $\alpha_{\rm DW,0}$, our models dissipate before $2\,{\rm Myr}$. In both the viscous and MHD-wind case, gas disk sizes are larger for larger initial masses. This is because in more massive disks, due to the larger column densities, CO can efficiently self-shield from UV-radiation further out \citep{Trapman23}.

The green tracks in Fig.~\ref{figure:models} display some representative gas disk size contours (for the smallest, average, and largest disks in our sample). Models with $R_{\rm c}>10\,{\rm au}$ over-predict the radial extent or our targets, both in the viscous and MHD-wind case (with no remarkable difference when models are evolved to only 1 or up to $3\,{\rm Myr}$), except for prohibitively low initial disk masses. Instead, initially more compact models ($R_{\rm c}=5\,{\rm au}$ or lower) can reproduce the observations essentially for $M_0\approx10^{-4}-10^{-5}\,M_\odot$ and $\alpha_{\rm SS}<10^{-4}$ for the viscous case. Such initial disk masses are at the edge of what is permitted by our thermochemical models to keep $^{13}$CO optically thick (cf., Fig.~\ref{figure:luminosities_plot}). In the MHD wind case, instead, it is relatively easier to explain such compact sources as long as $M_0\leq10^{-3}\,M_\odot$.

The bottom panel displays how these same models change when a moderate carbon depletion (by a factor of 10, as recently inferred by \citet{Trapman25_masses} in the sub-sample of Lupus disks targeted by the AGE-PRO ALMA Large Program, \citealt{Zhang25,Deng25}). Our models display naturally smaller sizes because of the lower CO column densities (and under the assumption that the faint optically thin CO tail predicted by thermochemical models, e.g., \citealt{Trapman22}, would still be undetectable at our deeper sensitivity). As a consequence, larger regions of the parameter space are compatible with our measurements, making it more viable to reproduce our observations also in the viscous case. 

In summary, viscous models are compatible with our observations only for extremely low initial disc masses and turbulence parameters, while MHD wind ones provide more flexibility, especially in the presence of moderate CO depletion. This being said, disc gas sizes alone are not able to conclusively exclude any of the two scenarios by themselves. Estimating gas masses (and, connected to them, CO depletion levels) for these targets could provide more robust constraints on their evolution mechanisms by breaking the degeneracies among the different scenarios discussed above. While this might be attempted by combining rare CO isotopologues and N$_2$H$^+$, as pioneered by \citet{Anderson19,Anderson22,Trapman22}, if feasible this would naturally be extremely observationally expensive due to the intrinsic faintness of our sources. On the other hand, our results corroborate the claim by \citet{Miotello21} that more than 50\% of the Lupus population could be made of such compact and faint disks, thus stressing how crucial these measurements are to understand disk evolution at a population level.

As a final remark, we stress that it was proposed that Lupus discs might be affected by late-infall of material from their parental cloud \citep{Winter24_lupus}. We speculate that, by replenishing our targets with fresh material, late-infall would play a double game. On the one hand it could help increase the lifetime of our compact models (supporting the validity of disk models with low initial masses in Fig.~\ref{figure:models}) to reproduce our observations. On the other hand, however, it would also lead to naturally larger disk sizes \citep[e.g.,][]{Kuffmeier23,Winter24,Padoan25}, exacerbating the necessity of initially small disk sizes, especially for our viscous models.

\subsection{Dust-to-gas size ratio}
A further diagnostic of disk evolution is the gas-to-dust size ratio $R_{\rm CO}/R_{\rm dust}$ \citep[e.g.,][]{Facchini17, Trapman19, Toci21}. If dust grains grow and drift inwards, the millimeter continuum emission becomes more compact than the gas thus increasing $R_{\rm CO}/R_{\rm dust}$, while dust traps can maintain more extended dust distributions with lower gas-to-dust size ratios. However, interpreting these ratios requires caution, since the optical depth of both gas and dust emission also plays a role.  In fact, \citet{Trapman19} showed that only size ratios $\gtrsim4$ can provide robust evidence of dust evolution.

In Lupus, previous analyses reported ratios ranging from 2 to 6 \citep[e.g.,][]{Ansdell18, Sanchis21, Long22}. Our sample shows a size ratio between 1 and 3, which is not indicative of strong radial drift. Furthermore, the lack of correlation between $R_{\rm CO}/R_{\rm dust}$ and $L_{\star}/M_{\star}$ further supports the idea that these objects are not dominated by drift, and may instead harbor unresolved small-scale substructures \citep[e.g.,][]{Toci21}. Alternatively, in line with the scenario proposed by \citet{Williams24}, our disks might simply have very high gas surface densities, such that millimeter-sized dust grains remain tightly coupled to the gas and do not drift significantly even in the absence of substructures. In either case, despite being radially compact, our disks display gas-to-dust ratios that are typical of the broader Lupus population. This suggests that the mechanisms that regulate dust–gas coupling operate in a similar way in both compact and extended systems. 

\section{Summary and conclusion}\label{section6}
We presented ALMA Band 7 observations of 17 faint protoplanetary disks in the Lupus star-forming region, targeting $^{12}$CO $(J=3-2)$ and $^{13}$CO $(J=3-2)$ emission. The data were analyzed using GoFish stacking techniques, image-plane flux extraction, and Gaussian fitting, enabling us to recover integrated CO fluxes and estimate gas and dust radii. 

The main conclusions from this work are:
\begin{enumerate}
    \item Of the 17 disks in our sample, 14 are detected in $^{12}$CO and 10 in $^{13}$CO. Among the $^{12}$CO detections, 7 sources are resolved, 5 are partially resolved, and 2 remain unresolved. Our average gas radius is $23.62\pm2.65\,{\rm au}$;
    \item Our observations, when compared with the models presented by \cite{Miotello21}, show disks compatible with optically thick $^{12}$CO and $^{13}$CO emission and intrinsically small gas and dust structures, supporting their interpretation that faint disks are primarily spatially compact;
    \item The derived gas radii are $\leq$ 40 au and the dust radii $\leq$ 20 au;
    \item Our gas-to-dust size ratio ranges between 1 and 3, consistent with the average value found so far in Lupus. This indicates no robust evidence for radial drift in our sample, a result further supported by the lack of correlation between the size ratio and the dust and gas sizes, as well as between the size ratio and $L_{\star}/M_{\star}$;
    \item As expected for optically thick $^{12}$CO emission, there is a monotonic positive correlation between $^{12}$CO luminosity and $^{12}$CO radius;
    \item We do not detect any correlation among the size ratio and the dust and gas sizes, the stellar masses and the disk masses, suggesting that, aside from extreme cases, most Lupus disks evolve in a similar way.
    \item Viscous evolution models require prohibitively small initial disc masses and turbulence level to be compatible with our small measured gas disc sizes, while MHD-wind models are generally more flexible in reproducing the observations. Independent estimates of the gas disc masses can provide more conclusive constraints on the mechanisms regulating the evolution of these sources, the bulk of Lupus population.
\end{enumerate}

\begin{acknowledgements}
This article makes use of the following ALMA data: ADS/JAO.ALMA\#2023.1.00428.S. ALMA is a partnership of the ESO (representing its member states), the NSF (USA) and NINS (Japan), together with the NRC (Canada) and the NSC and ASIAA (Taiwan), in cooperation with the Republic of Chile. The Joint ALMA Observatory is operated by the ESO, the AUI/NRAO, and the NAOJ. CFM is funded by the European Union (ERC, WANDA, 101039452). Views and opinions expressed are however those of the author(s) only and do not necessarily reflect those of the European Union or the European Research Council Executive Agency. Neither the European Union nor the granting authority can be held responsible for them. GR is funded by DiscEvol (ERC Starting Grant, grant agreement No. 101039651) and acknowledges additional support from Fondazione Cariplo (grant No. 2022-1217). SF is funded by UNVEIL (grant agreement No. 101076613) and acknowledges financial support from PRIN-MUR (grant No. 2022YP5ACE). Views and opinions expressed are those of the authors only and do not necessarily reflect those of the European Union, the European Research Council or other funding agencies. Neither the European Union nor the granting authorities can be held responsible for them.
\end{acknowledgements}

\bibliographystyle{aa}
\bibliography{biblio}

\begin{appendix}
\onecolumn
\section{GALARIO uv-plots}
In this appendix, we present the collection of visibility ($uv$) plots for the continuum of the whole sample, along with a corner plot of the target Sz~131.

\begin{figure*}[h!]
    \centering
    \includegraphics[width=\textwidth]{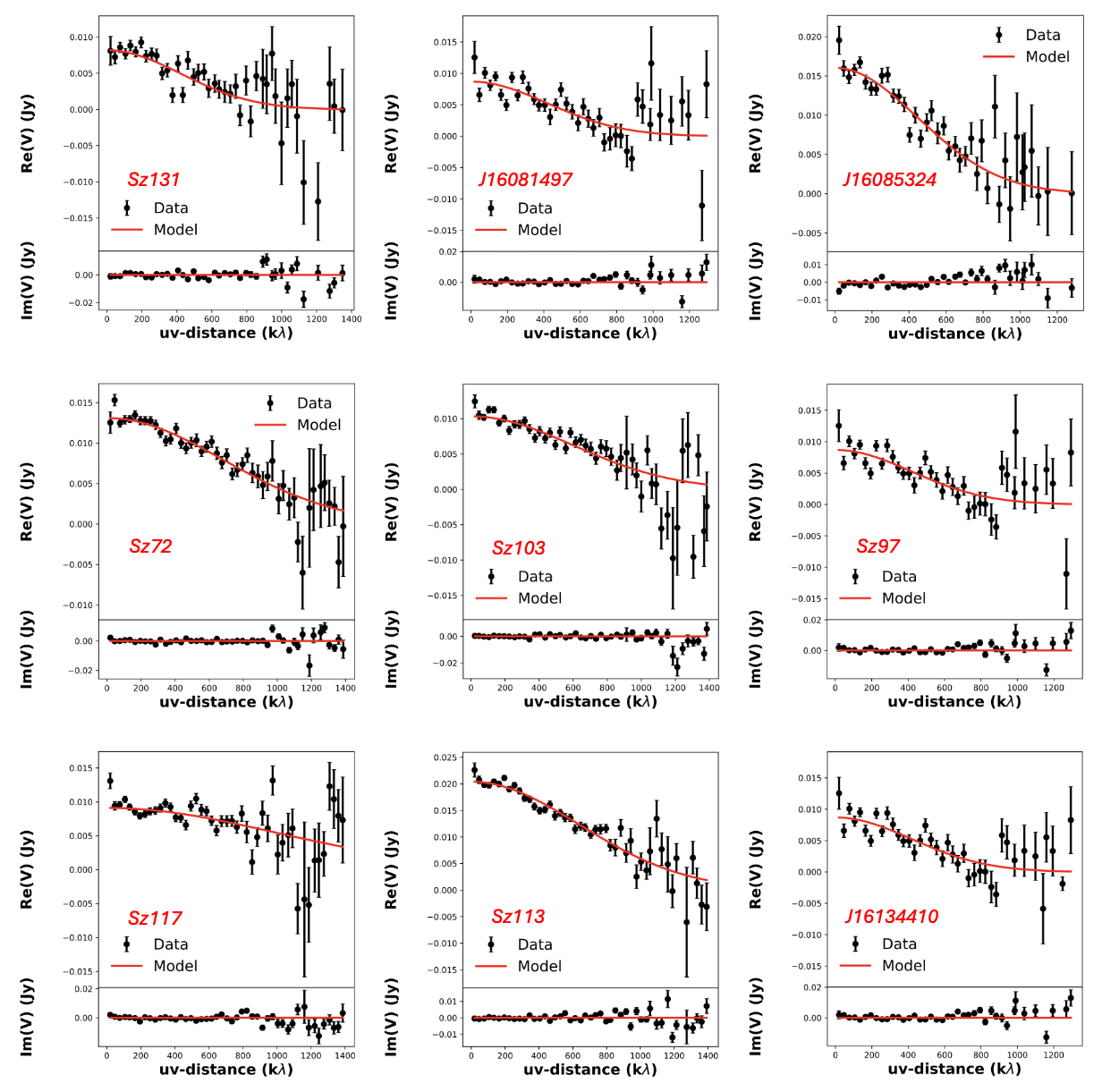}
    \caption{Visibility plots of observed and modeled visibilities, obtained with \texttt{GALARIO} by fitting a Gaussian profile to the millimeter emission.}
    \label{figure:uvplots}
\end{figure*}

\begin{figure*}[h!]
    \centering
    \addtocounter{figure}{-1}
    \includegraphics[width=\textwidth]{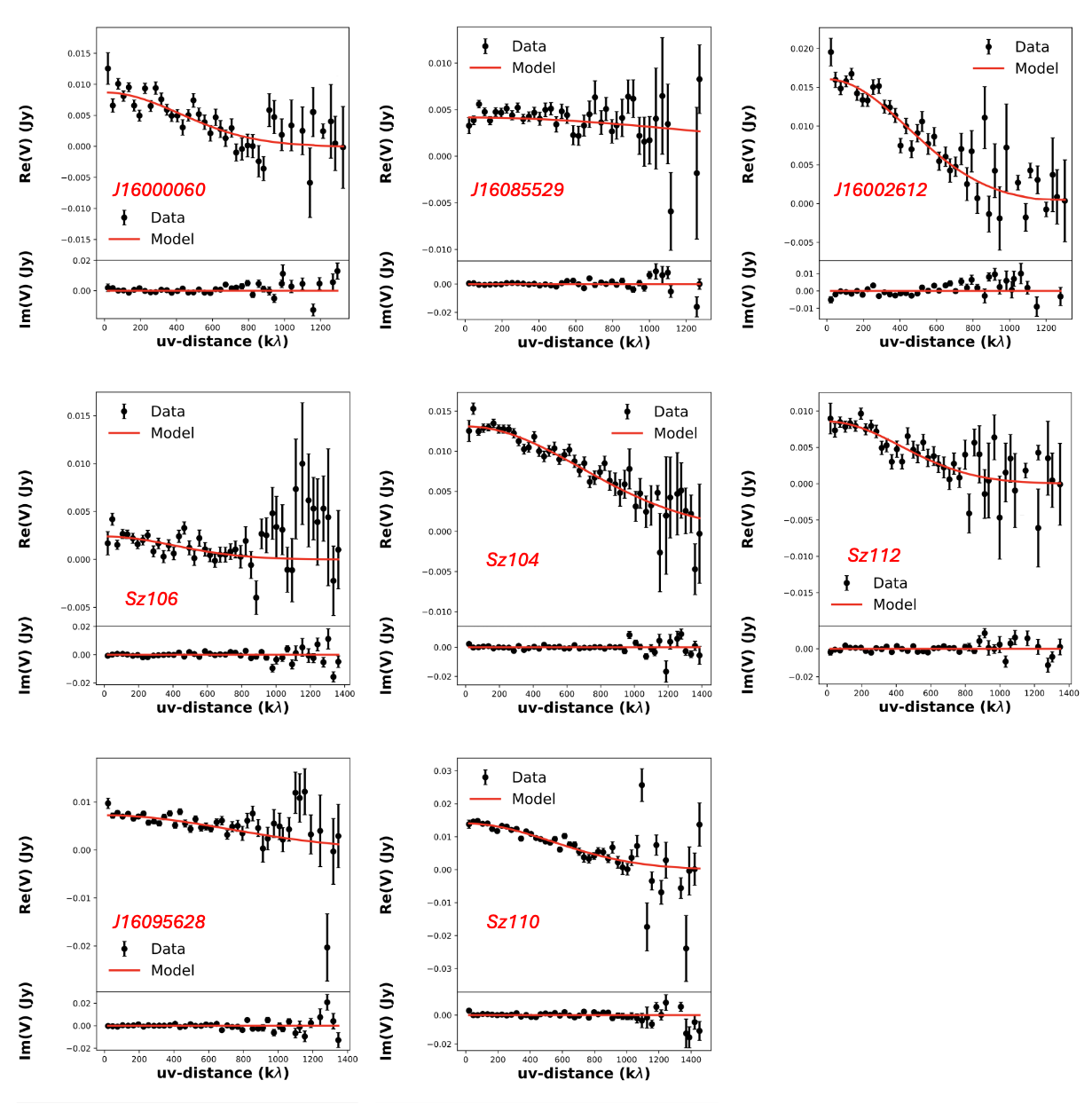}
    \caption{Fig.~\ref{figure:uvplots} (continued).}
\end{figure*}

\begin{figure*}[h!]
    \centering
    \includegraphics[width=\textwidth]{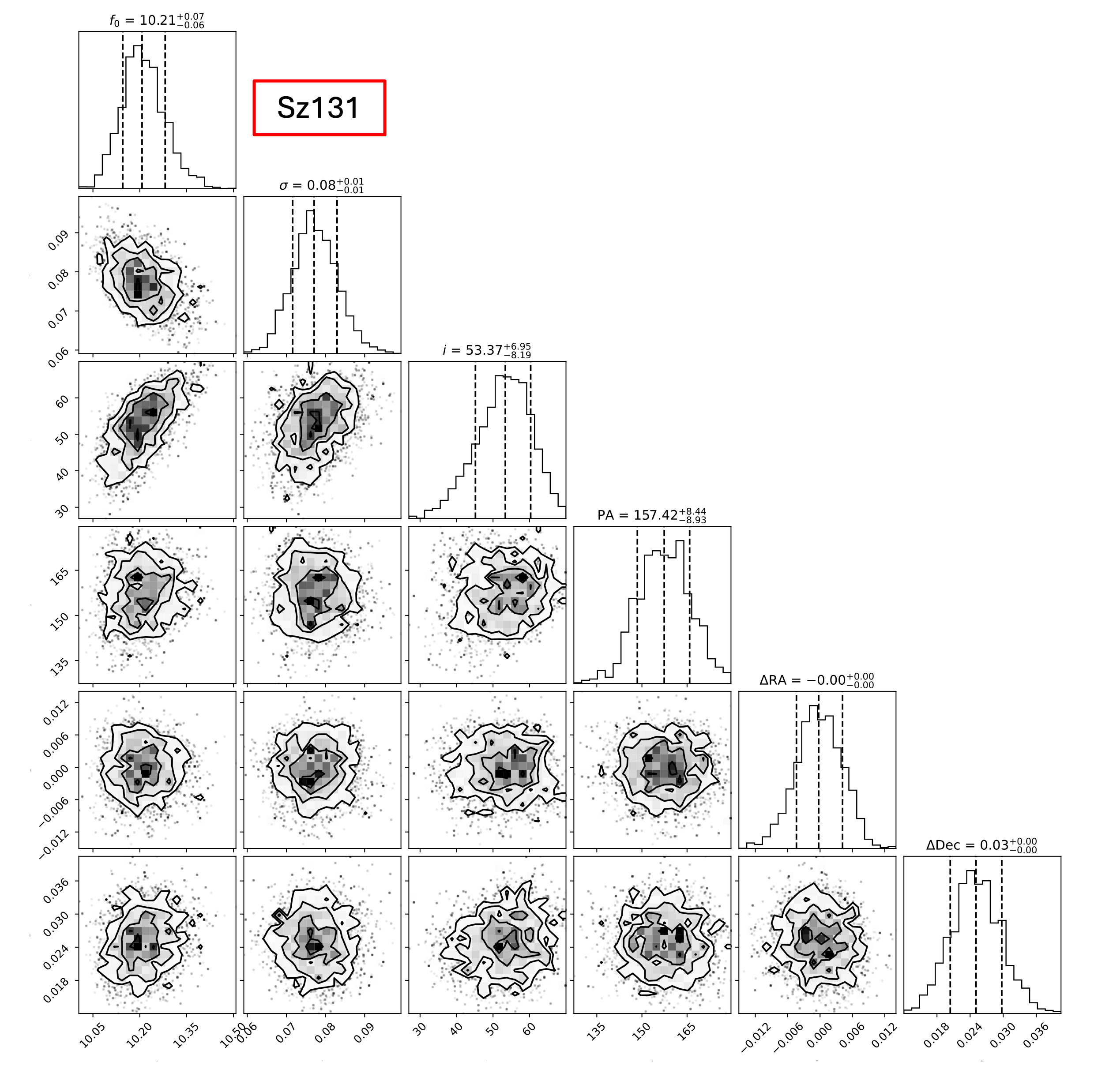}
    \caption{Corner plot for Sz~131 displaying the marginalized posterior distributions of the free parameters obtained from the MCMC fit. Dashed vertical lines indicate the median values $\pm$ 1$\sigma$ confidence intervals.}
\end{figure*}

\FloatBarrier
\section{Size ratio against stellar proprieties}
\begin{figure*}[h!]
    \centering
    \includegraphics[width=\textwidth]{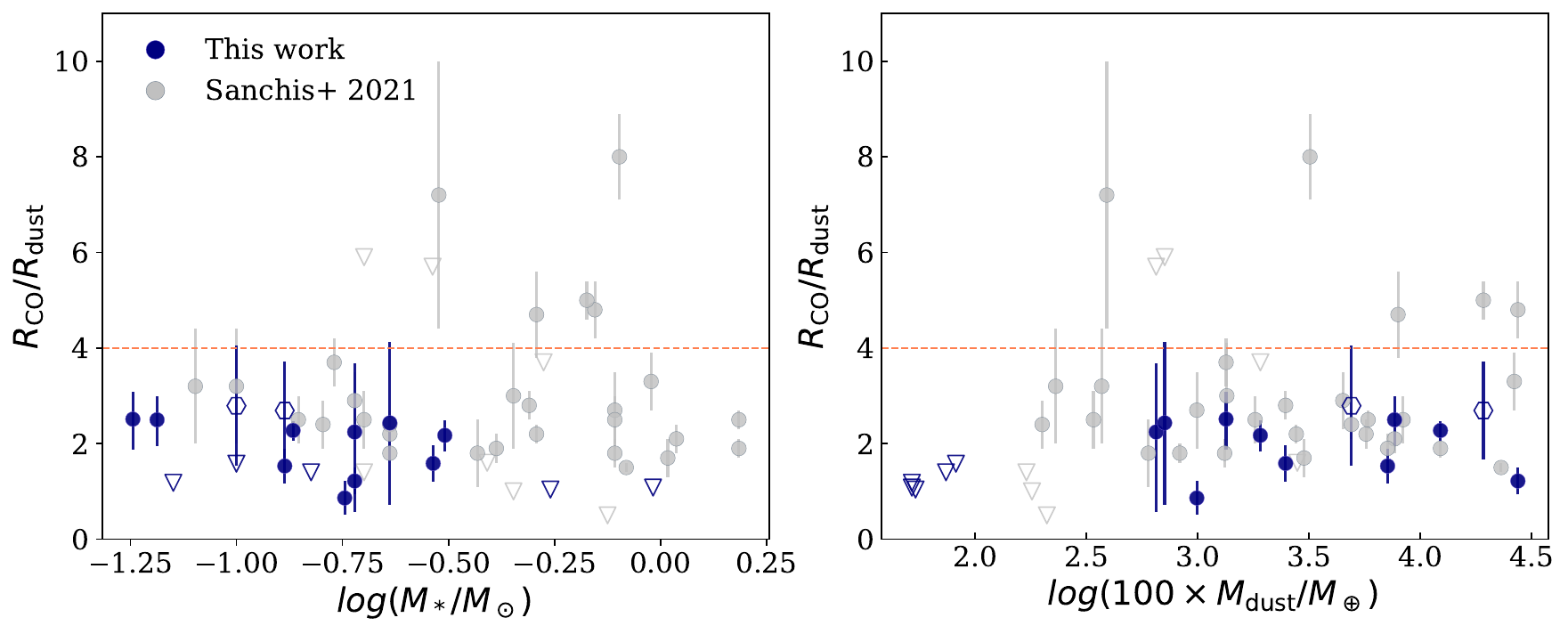}
    \caption{Size ratio as a function of the some stellar proprieties. The results of this work are plotted as blue dots, those by \cite{Sanchis21} as gray dots. The upper limits are indicated as triangles.}
    \label{figure:ratio_vs_stellar}
\end{figure*}

In this section, we present some additional analyses of the size ratio. In particular, Figure \ref{figure:ratio_vs_stellar} shows the size ratio as a function of stellar mass and disk mass. No correlation is observed between these quantities. By combining our sample with the disk population from \citet{Sanchis21}, we cover a wide range of stellar and disk masses, as well as dust and CO sizes. Overall, the disks exhibit a broadly consistent trend, with only a few exceptions in the gray sample, indicating that the majority of the disk population behaves in a similar manner, largely independent of stellar or disk properties. 

\FloatBarrier
\section{Table and Observing logs}

\begin{table*}[hbt!]
\caption{Stellar properties and coordinates of the targets included in our sample.} 
\label{table:stellar_par}
\centering  
\begin{tabular}{l l l l l} 
\hline\hline 
Source & RA & Dec & Stellar Mass & $d$ \\
     & (J2000) & (J2000) & ($M_\odot$) & (pc) \\
\hline\hline 
        Sz~131 & 16:00:49.414 & $-$41:30:04.263 & 0.31 & 160.6 \\
        J15450634-3417378~$^{1}$ & 15:45:06.322 & $-$34:17:38.332 & -- & 158.0 \\
        J16081497-3857145 & 16:08:14.960 & $-$38:57:14.910 & 0.06 & 158.8 \\
        J16085324-3914401 & 16:08:53.227 & $-$39:14:40.553 & 0.29 & 163.0 \\
        Sz~72 & 15:47:50.608 & $-$35:28:35.779 & 0.14 & 160.5 \\
        Sz~103 & 16:08:30.254 & $-$39:06:11.538 & 0.19 & 157.1 \\
        Sz~97 & 16:08:21.787 & $-$39:04:21.860 & 0.19 & 157.3 \\
        Sz~117 & 16:09:44.348 & $-$39:13:30.555 & 0.23 & 156.9 \\
        Sz~113 & 16:08:57.786 & $-$39:02:23.260 & 0.13 & 160.5 \\
        J16134410-3736462 & 16:13:44.082 & $-$37:36:46.598 & 0.96 & 158.5 \\ 
        J16000060-4221567 & 16:00:00.584 & $-$42:21:57.203 & 0.15 & 159.4 \\
        J16085529-3848481 & 16:08:55.275 & $-$38:48:48.551 & 0.07 & 155.6 \\ 
        J16002612-4153553 & 16:00:26.099 & $-$41:53:55.782 & 0.1 & 163.2 \\
        Sz~106 & 16:08:39.743 & $-$39:06:25.701 & 0.5 & 158.7 \\
        Sz~104 & 16:08:30.798 & $-$39:05:49.238 & 0.1 & 159.8 \\
        Sz~112 & 16:08:55.513 & $-$39:02:34.359 & 0.1 & 159.3 \\ 
        J16095628-3859518 & 16:09:56.281 & $-$38:59:51.973 & 0.06 & 157.1 \\
        Sz~110 & 16:08:51.553 & $-$39:03:18.087 & 0.2 & 157.5 \\ 
\hline
\end{tabular}

\begin{tablenotes}
\small
\item These values are taken from \cite{Manara23, Alcala14, Alcala17}. The uncertainties on stellar masses are 0.1 dex \citep{Alcala17}.  \\
$^{1}$ this source isn't included in the subsequent analysis.
\end{tablenotes}

\end{table*}

\begin{table*}[h!]
\caption{Spectral Windows of the Spectral Settings of the ALMA Band 7 Observations.}
\label{table:setup}
\centering
\begin{tabular}{l l l l l l}
    \hline\hline
    Baseband & Central rest freq. & SPW name & $\#$ Channels & Bandwidth & Resolution \\
    & (GHz) &  &  & (MHz) & (km/s) \\
    \hline\hline 
    1 & 345.795990 & CO $v=0,\, J=3-2$. & 1920 & 58.59 & 0.071 \\
    1 & 344.200109 & HC$^{15}$N $v=0,\, J=4-3$. & 1920 & 58.59 & 0.071 \\
    2 & 342.882857 & CS $v=0,\, J=7-6$. & 1920 & 117.19 & 0.171 \\
    3 & 330.587965 & $^{13}$CO $v=0,\, J=3-2$. & 960 & 58.59 & 0.177 \\
    4 & 332.900000 & Continuum & 3840 & 1875.00 & 1.407 \\
    \hline
\end{tabular}
\end{table*}

\begin{table*}[h!]
\caption{Observational setup and conditions for ALMA Band 7 execution blocks.}
\label{table:conditions}
\centering
\begin{tabular}{l l l l l l}
    \hline\hline
    Execution Block & Group & $\#$ Antennas & Integration Time & PWV & Phase RMS \\
    &  &  & (min) & (mm) & ($\mu$m) \\
    \hline\hline 
    X118bf6a/X3c09 & 1 & 44 & 7.67 & 1.07 & 55.854 \\
    X11adad7/X316a9 & 2 & 45 & 47.55 & 0.83 & 55.328 \\
    X11b9826/X11fe7 & 2 & 43 & 47.53 & 0.48 & 55.026 \\
    X11b9826/X1187c & 2 & 43 & 47.55 & 0.56 & 50.225 \\
    X11b9826/X1187c & 2 & 42 & 47.62 & 0.56 & 38.249 \\
    \hline
\end{tabular}
\end{table*}

\thispagestyle{empty}

\begin{table}[hbt!]
\centering
\caption{Fluxes, radii and geometry of the targets included in our sample.}
\label{table:sample}
\begin{adjustbox}{angle=90,left}   
  \resizebox{0.95\textheight}{!}{
    \begin{tabular}{l l l l l l l l l l}
    \hline\hline
    Source & F$_{\rm 12_{CO}}$ & F$_{\rm 13_{CO}}$ & R$_{\rm 68, CO}$ & R$_{\rm 90, CO}$ & R$_{\rm 68, dust}$ & R$_{\rm 90, dust}$ & $R_{\rm CO}/R_{\rm dust}$ & $i$ & PA 
    \\
     & (mJy~km~s$^{-1}$) & (mJy~km~s$^{-1}$) & (au) & (au) & (au) & (au) &  & (deg) & (deg)
    \\
    \hline\hline 
        \textbf{\textit{Shift and stacked line emission}} \\
        \vspace{0.2cm}
        Sz131 & 2442.28$\pm$353.72 & 1688.48$\pm$356.13 & 40.9$^{+4.9}_{-4.9}$ & 58.1$^{+6.9}_{-6.9}$ & 18.82$^{+1.93}_{-1.61}$ & 26.82$^{+2.72}_{-2.25}$ & 2.17$^{+0.48}_{-0.43}$ & 47.28 & 335.0 \\
        \vspace{0.2cm}
        J16081497-3857145 & 1689.51$\pm$294.97 & 1144.10$\pm$347.47 & 44.5$^{+7.6}_{-7.6}$ & 63.2$^{+10.79}_{-10.79}$ & 17.81$^{+2.41}_{-1.66}$ & 25.33$^{+3.47}_{-2.41}$ & 2.49$^{+0.73}_{-0.67}$ & 58.24 & 80.13 \\
        \vspace{0.2cm}
        J16085324-3914401 & 2863.89$\pm$385.07 & 2166.74$\pm$404.15 & 28.7$^{+6.7}_{-6.7}$ & 40.7$^{+9.5}_{-9.5}$ & 18.09$^{+0.82}_{-0.82}$ & 25.75$^{+1.30}_{-1.14}$ & 1.58$^{+0.46}_{-0.43}$ & 33.0 & 100.18 \\
        \vspace{0.2cm}
        Sz72 & 2694.85$\pm$169.94 & 1710.48$\pm$181.30 & 26.3$^{+1.9}_{-1.9}$ & 37.34$^{+2.7}_{-2.7}$ & 11.56$^{+0.64}_{-0.48}$ & 16.55$^{+0.96}_{-0.80}$ & 2.26$^{+0.29}_{-0.28}$ & 31.46 & 217.99 \\
        \vspace{0.2cm}
        Sz103 & 1405.71$\pm$205.22 & 863.26$\pm$204.22 & 15.9$^{+3.6}_{-3.6}$ & 22.6$^{+5.1}_{-5.1}$ & 13.05$^{+0.63}_{-0.63}$ & 18.56$^{+0.94}_{-0.79}$ & 1.22$^{+0.34}_{-0.32}$ & 51.13 & 260.87 \\
        \vspace{0.2cm}
        Sz97 & 1215.64$\pm$236.30 & $<$718.39 & 12.7$^{+6.7}_{-6.7}$ & 18.0$^{+9.5}_{-9.5}$ & 5.66$^{+2.99}_{-2.05}$ & 8.03$^{+4.25}_{-2.83}$ & 2.24$^{+3.04}_{-1.05}$ & 45.0 & 61.13 \\
        \vspace{0.2cm}
        Sz117 & 942.24$\pm$234.66 & 697.55$\pm$211.56 & 19.1$^{+13.2}_{-13.2}$ & 27.1$^{+18.7}_{-18.7}$ & 7.85$^{+0.94}_{-0.78}$ & 11.13$^{+1.26}_{-1.10}$ & 2.43$^{+2.13}_{-1.76}$ & 40.05 & 95.22 \\
        \vspace{0.2cm}
        Sz113 & 3423.67$\pm$285.09 & 1983.61$\pm$320.72 & 19.2$^{+4.5}_{-4.5}$ & 27.3$^{+6.4}_{-6.4}$ & 12.53$^{+0.32}_{-0.32}$ & 17.66$^{+0.48}_{-0.48}$ & 1.55$^{+0.42}_{-0.39}$ & 5.0 & 116.43 \\
        \vspace{0.2cm} \\
        \hline
        \textbf{\textit{Flux extraction from the integrated intensity maps}} \\
        \vspace{0.2cm}
        J16134410-3736462 & $<$5.60 & $<$11.56 & $<$16.48 & $<$23.40 & 15.32$^{+0.01}_{-0.32}$ & 19.13$^{+0.02}_{-0.16}$ & $<$1.08 & 49.40 & 69.90 \\
        \vspace{0.2cm}
        J16000060-4221567 & 37.75$\pm$6.14 & $<$8.75 & $<$14.70 & $<$20.87 & 10.52$^{+5.99}_{-3.55}$ & 15.00$^{+5.89}_{-4.78}$ & $<$1.4 & 65.7 & 86.95\\
        \vspace{0.2cm}
        J16085529-3848481 & $<$15.94 & $<$7.87 & $<$19.44 & $<$27.61 & 16.49$^{+0.93}_{-3.07}$ & 26.76$^{+0.31}_{-5.89}$ & $<$1.18 & 68.0 & 88.3\\ 
        \vspace{0.2cm}
        J16002612-4153553 & 18.27$\pm$4.11 & $<$7.81 & 33.7$^{+15.2}_{-15.2}$ & 47.9$^{+21.6}_{-21.6}$ & 12.07$^{+0.33}_{-0.49}$ & 15.65$^{+0.16}_{-0.65}$ & 3.06$^{+1.57}_{-1.39}$ & 53.11 & 167.56 \\
        \vspace{0.2cm}
        Sz106 & $<$0.01 & $<$0.09 & $<$19.81 & $<$28.13 & 19.04$^{+5.71}_{-3.65}$ & 26.98$^{+8.25}_{-5.08}$ & $<$1.04 & 81.31 & 78.7 \\
        \vspace{0.2cm}
        Sz104 & 20.78$\pm$5.17 & 20.65$\pm$5.43 & $<$13.9 & $<$19.74 & 8.79$^{+4.79}_{-2.24}$ & 12.65$^{+6.72}_{-3.20}$ & $<$1.58 & 31.0 & 66.1\\
        \vspace{0.2cm}
        Sz112 & 25.42$\pm$4.24 & $<$14.19 & 28.3$^{+9.2}_{-9.2}$ & 40.2$^{+13.0}_{-13.0}$ & 10.51$^{+2.07}_{-2.07}$ & 14.98$^{+2.87}_{-2.87}$ & 2.68$^{+1.71}_{-1.16}$ & 31.46 & 47.89 \\ 
        \vspace{0.2cm}
        J16095628-3859518 & 126.96$\pm$15.37 & 24.12$\pm$6.60 & 21.3$^{+4.0}_{-4.0}$ & 30.2$^{+5.7}_{-5.7}$ & 8.48$^{+1.41}_{-1.10}$ & 12.10$^{+2.36}_{-1.41}$ & 2.50$^{+0.86}_{-0.80}$ & 30.0 & 87.93 \\
        \vspace{0.2cm}
        Sz110 & 215.25$\pm$11.91 & 61.48$\pm$6.49 & 12.5$^{+5.1}_{-5.1}$ & 17.8$^{+7.2}_{-7.2}$ & 14.49$^{+0.47}_{-0.47}$ & 20.65$^{+0.63}_{-0.63}$ & 0.86$^{+0.39}_{-0.36}$ & 49.55 & 13.10 \\ 
        \vspace{0.2cm} \\
    \hline
    \end{tabular}
    }
\end{adjustbox}

\begin{tablenotes}
\small
\item $^{12}$CO and $^{13}$CO fluxes, $^{12}$CO and dust radii, size ratio and geometry for the sources in our sample of the faint disks in Lupus. 
\end{tablenotes}

\end{table}

\end{appendix}
\end{document}